\RequirePackage[undo-recent-deprecations]{expl3}
\documentclass[a4paper,review,fleqn]{cas-sc}

\usepackage[numbers]{natbib}
\usepackage{lineno}
\usepackage{textcomp}
\usepackage{anyfontsize}
\usepackage{graphicx}
\usepackage{graphics}
\usepackage{geometry}
\usepackage{caption}
\usepackage{subcaption}
\usepackage{adjustbox,lipsum}
\usepackage{float}
\usepackage{amssymb}
\usepackage{amsmath}
\usepackage{color}
\usepackage{xcolor}  
\usepackage{hyperref}  
\hypersetup{colorlinks,urlcolor=blue,linkcolor=blue,citecolor=blue,filecolor=blue,urlcolor=blue}
\usepackage{multirow}
\usepackage{longtable}
\usepackage{lscape}
\newtheorem{definition}{Definition}

\begin{document}
  \let\WriteBookmarks\relax
  \def\floatpagepagefraction{1}
  \def\textpagefraction{.001}
  \shorttitle{Non-extensive realizations in interacting ion channels}
  \shortauthors{D O C Santos et~al.}


\title[mode = title]{Nonextensive realizations in interacting ion channels: implications for 
mechano-electrical transducer mechanisms}                      

\author[1]{D O C Santos}
\address[1]{Universidade Federal do Sul da Bahia, CEP $45600-923$, Itabuna, Bahia, Brazil}
\credit{Computational simulations, Data analysis, Original draft preparation, Writing}

\author[2]{M A S Trindade}
\address[2]{Departamento de Ci\^encias Exatas e da Terra, Colegiado de Física, Universidade do Estado da Bahia, CEP $41150-000$, Salvador, Bahia, Brazil}
\credit{Theoretical calculations}

\author[1]{A J da Silva}
\cormark[1]
\ead{adjesbr@ufsb.edu.br,adjesbr@gmail.com}
\credit{Conceived the study, Data analysis, Original draft preparation, Writing}
\cortext[cor1]{Corresponding author}

\begin{abstract}
Although there are theoretical studies on the thermodynamics of ion channels, an investigation involving the thermodynamics of coupled channels has not been proposed. To overcome this issue, we developed calculations to present a thermodynamic scenario associated with mechanoelectrical transduction channels as a single and coupling of two-state channels. The modeling was inspired by the Tsallis theory, in which we derived the open and closed probability distributions, the joint probability distribution, the Tsallis entropy, and the Shannon mutual information. Despite being well studied in many biological systems, the literature has not addressed both entropy and mutual information related to isolated and a pair of physically interacting mechanoelectrical transduction channels. Inspired by the hair cell biophysics, we revealed how the presence of nonextensivity modulates the degree of entropy and mutual information as a function of stereocilia displacements. In this sense, we showed how the non-extensivity regulates the current versus displacement curve for a single and two interacting channels made up of a single open and closed states. Overall, subadditivity and superadditivity yielded increments and decrements in the entropy and mutual information compared with the extensive regime. We also observed that the magnitude of the interaction between the two channels significantly influences the amplitude of the joint entropy and the mutual information. These results are directly related to the modulation of the channel kinetics, given by changes evoked by hair cell displacements. Finally, we found that the gating force modulates the contribution of subadditivity and superadditivity present in the joint entropy and the mutual information. The present findings shed light on the thermodynamic process involved in the molecular mechanisms of the auditory system. 
\end{abstract}


\maketitle  

\section{Introduction}

Knowing how biological systems perform their operations provides an understanding of the mechanisms behind both normal and pathological physiological conditions. Biologically inspired models constitute a tool for approximating theoretical predictions related to experimental work. The extensive development of analytic and numerical methods translating physical laws on which the biological system is dependent is helping to provide accurate descriptions of experimental results. In addition, a theoretical study offers the possibility of simulating inaccessible experimental protocols. Among the theoretical frameworks used to comprehend biological systems, statistical physics allows assessing different classes of phenomena (\cite{stanley1994}). In this context, Boltzmann statistical theory has been applied to quantifying and uncovering essential aspects of ion channel dynamics. In their interesting review, Dubois \textit{et al.} stated: "\textit{Ion channels control a great number of cellular functions including electrical activity, hormone release, cell proliferation, apoptosis, and migration. They are activated or inactivated by different stimuli and statistically behave following the notion of probability introduced by Ludwig Boltzmann. Indeed, since the end of the 19th century, it is accepted that an atom, an ion, or a molecule is not in a stable state but has a defined probability of being in a given state}" (\cite{Dubois2009}). Nevertheless, despite its unquestionable importance and usefulness, the probabilistic model developed by Boltzmann presents limitations when physical interactions or long-range correlations are present in the system. To overcome this problem, non-extensive statistics or generalized thermostatistics $(GTS)$ has been receiving increasing attention. Developed by Tsallis at the end of the 1980s, $GTS$ describes long-range correlations or interactions in complex systems such as those observed within the biological realm, describing many phenomena more accurately than using the Boltzmann framework. The calculation of entropy provides valuable information about a specific state, where higher entropy implies a more complex regime as compared to lower entropic values (\cite{Xiong2017}). Recently, Montufar \textit{et al.} studied the $RR$-interval time series from electrocardiogram recordings, reporting a reduced entropy in sedentary people concerning people who practice some physical activity \cite{SolisMontufar2020}). In a theoretical context, Suyari highlighted that successful descriptions and utilization in $GTS$ originates from the presence of self-similarity in the $q$-product. In this context, he applied the $q$-product rule to present important results in the following themes: the law of error, the $q$-Stirling's formula, the $q$-multinomial coefficient, and Pascal triangle (\cite{Suyari2006}).

As a typical example of complexity, physiological systems offer plenty of room for investigating non-extensive phenomena. For example, researchers recorded the existence of long-range correlations in electroencephalogram signals (\cite{Fang2010}). Furthermore, Silva \textit{et al.} presented a study of cellular communication and neuroplasticity assuming $GTS$ premises (\cite{Silva2018}). Also, this theory was applied to uncover non-random mechanisms in electrocardiograms and cellular motion in two-dimensional cellular aggregates (\cite{ACHARYA2018,UPADHYAYA2001}). Regarding ion channels, it was recently proposed that a model describing the packing of atoms around the pore of a voltage-gated sodium channel $(NavAb)$ shows non-extensivity in the inter-atomic hydropathic interactions around the pore (\cite{Xenakis2021}). Furthermore, it is worth mentioning that Erden firstly proposed a model describing the steady-state properties of a voltage-gated ion channel using the $GTS$ framework, recovering the extensive or classical properties through a particular limit of the non-extensive index (\cite{Erdem2007}). In summary, $GTS$ represents an attractive theory for modeling ionic channels in which physical interactions and long-range correlations may be investigated, considering the influence of different physiological parameters.

There is increasing evidence that ion channels, believed to act independently in channel clusters, may interact in several situations. For instance, Liu and Dilger proposed an Ising model to study the interaction between ion channels activated by ligands (\cite{Liu1993}). Also, Draber \textit{et al.} studied the cooperativity between potassium channels in the distribution of residence times, verifying the existence of a deviation from the binomial distribution for identical and independent channels (\cite{Draber1993}). In this context, Uteshev pointed out that the adoption of the binomial distribution, customarily applied to detect the presence of interactions between channels, has limitations (\cite{Uteshev1993}). Additionally, to study cooperative mechanisms, Markovian models were also proposed. In this framework, Keleshian \textit{et al.} developed a dependency model for a pair of identical channels, each modeled by a continuous-time Markov chain, where the transition rates depended on the other channel's conductance status (\cite{Keleshian1994}). Ball \textit{et al.} modeled both an isolated and a set of independent and identically distributed channels, described by continuous time-reversible Markov chains, in which a random medium was acting over a single channel or a cluster. This strategy allowed these researchers to obtain dependency relations between channels (\cite{Ball1994}). Subsequently, Dekker and Yellen combined patch-clamp experiments and noise analysis of macroscopic hyperpolarization-activated cyclic nucleotide-gated $(HCN)$ channel currents. They demonstrated using non-stationary fluctuation analysis the presence of cooperativity among them (\cite{Dekker2006}). Also, another study showed that clusters of voltage-gated calcium channels $(Cav1.3)$ functionally coupled surround clusters of calcium-activated potassium channels $(BK)$. According to the results, the most plausible mechanisms seem to be the increase in calcium availability through $Cav1.3$ channels, preventing $BK$ currents by activating these channels only close to the threshold potential (\cite{Vivas2017}). As described below, investigations in the auditory nervous system point to cooperative mechanisms between channels.

The physiological mechanisms of audition are not yet accurately known. However, the theoretical model developed by Lord Rayleigh at the beginning of the last century helped to shed light on this issue (\cite{LordRayleigh1907}). The human auditory system is constituted by three compartments: outer, middle, and inner ears. In particular, the inner ears are responsible for detecting and amplifying the intensity of different sound ranges. The hair cells within this region are the main detector element associated with the mechanoelectrical transducer machinery. They transform the sound vibrations into electrical signals transferred by the auditory nerve to the auditory brainstem and auditory cortex. The main element that performs such conversion is the hair bundle or stereocilia placed on the top of these cells. These mechanically sensitive organelles are extremely sensitive, responding to very low-intensity mechanical stimuli. Deflections of the hair bundle trigger the mechanoelectrical conversion via a pull on the tips of the stereocilia, opening and closing the mechanoelectrical $(MET)$ channels (\cite{Pickles2012}). Only in recent years have the auditive network's molecular mechanisms been uncovered. Thanks to electrophysiological, ultrastructural studies, and mathematical modeling, all physiological complexity involved in the transduction mechanism is gradually revealed. Recently, Gianolli \textit{et al.} developed a model of the hair-cell mechanoelectrical transduction cooperativity, considering $MET$ channels at the auditory sensory epithelium in the cochlea. In their modeling of cooperation between $MET$ channels, they used molecular kinetics directly linked to the oscillation of the hair bundle displacement. The tip-link connection governed both approximation and departure of two neighbor channels, also mediated by the lipid bilayer. Consequently, the system may be classified as a coupled mechanical oscillator. Their biomechanical formulation reproduced the main properties of hair-cell mechanoelectrical transduction (\cite{Gianoli2017,Gianoli2019}).

Different ion channels and pumps regulate cellular ion levels, keeping the synaptic transmission from the hair cell to other brain areas. Several authors have characterized the electrical pattern of the ionic regulation on nervous electrical activity. Although studies in human hair cells pointed to four $MET$ channel candidates, additional support for this possibility is still required. Furthermore, investigations employing transmission electron micrograph show two neighbor stereocilia physically connected with a tip link composed of cadherin filaments. At the end of each cell, there are $MET$ channels. This peculiar arrangement implies that both voltage and mechanical perturbations regulate channel kinetics where the kinetic of the first channel affects the dynamics of a second one and vice-versa. This is directly reflected in the gating scheme of a two-state interacting channel in which potential and displacement energies modulate the transition between closed and open states. In this substrate, the electrical response of hair cells arises in response to bundle displacements promoted by the mechanical stimulus. Boltzmann sigmoid equations were adopted to adjust the experimental results in this framework. Nevertheless, this assumption contradicts the experimental observations again since Boltzmann formalism does not consider interactions in its theoretical foundations. Indeed, Gianoli \textit{et al.} assumed the absence of interactions in the system, configuring an apparent contradiction with the molecular substrate used in the theoretical construction.

Regarding the analysis of cooperative $MET$ channels, another important problem is to quantify the system information and entropy since interactions imply mutual changes in channel kinetics. The $GTS$ formalism might provide an answer since it proposes understanding how long-range correlations modulate entropy and information. Regarding the analysis of cooperative channels, the calculation of entropy and information is currently being made only in the sense of classical statics physics. However, studies involving isolated ion channels investigated the entropy, but situations involving cooperativity between channels rarely are reported in the literature. For instance, Duran and Marzen studied mutual information from the response of voltage-activated potassium channel $(Kv)$ when the potassium concentration is modified (\cite{Duran2020}). In addition, Lewin \textit{et al.} studied entropy-based modulation of the $Kv$ cluster (\cite{Lewin2020}), while Portella \textit{et al.} investigated the effect of channel dependence on potassium ion permeation, concluding that entropy represents the major contribution to the permeation through the channel (\cite{Portella2008}). Relative to the information theory, Hichri \textit{et al.} quantified voltage-activated sodium channels $(NaV1.5)$ clustering using Shannon’s entropy (\cite{Hichri2020}). Moreover, Gatenby and Frieden applied information theory to study the ionic traffic, suggesting that transmembrane ion gradients and propagated action potentials in neurons are also important for non-neuronal cells surviving (\cite{Gatenby2017}). Despite the importance of these studies, a theoretical construction that explicitly describes the physical interaction between channels is still lacking. In this sense, $GTS$ represents an appropriate theory, suitable to quantify both entropy and existing information in interacting systems such as $MET$ channels.

In this work, we provide a theoretical formulation, showing how the presence of non-extensivity may be crucial for an appropriate description of ion channel cooperation. A generalized formulation will be applied to a kinetic model of two channels with single open and closed states in this context. We finally calculate kinetics, the Tsallis entropy, and mutual information due to single and double cooperative $MET$ channels.  

\section{Model design}

In this section, we start by presenting a formulation involving joint probability distributions, the $q$-product, and the Tsallis entropy. This theoretical approach will be applied to analyze interactions between two $MET$ channels in terms of a generalized entropy. The $q$-product introduces correlations between distributions of random variables so that in the limit $q \rightarrow 1$, the joint probability distribution is the product of probability distributions and, therefore, leads to independence between random variables distributions(\cite{Borges2004,Nivanen2003}). Finally, we apply the theoretical construction to the case of two $MET$ channels described by non-extensive steady-state probabilities, where the $q$-product governs the magnitude of channel cooperativity.

\subsection{Joint probability distributions based on the $q$-product and Tsallis entropy}

When we consider a system composed of two independent continuous-time, discrete random processes, characterized by the probability distributions, $p(x)$ and $p(y)$, the joint probability of the composed system is simply the product of the probability distributions, $p(x,y)= p(x)p(y)$. In this sense, let us define a generalization of the joint probability distribution to introduce correlations between the random processes: 
\begin{equation}
p(x,y) := \frac{p(x) \otimes_{q} p(y)}{\sum_{x \in \tilde{\mathcal{X}}} \sum_{y \in \tilde{\mathcal{Y}}}[p(x) \otimes_{q} p(y)]},
\label{pxyqproduto}
\end{equation}
where $\otimes_{q}$, denotes the $q$-product and $q$ represents the entropic index in the nonextensive formalism (\cite{Borges2004,Suyari2006,Tsallis2009}). For numbers $a$ and $b$, where $a \geq 0, b \geq 0$:
\begin{equation}
a \otimes_{q} b := [a^{1-q}+b^{1-q}-1]_{+}^{\frac{1}{1-q}},
\end{equation}
in which $[A]_{+}:= max(0,A)$ and in the limit $q \rightarrow 1$ we recover the usual product.

Notice that this joint probability is commutative: $p(x,y) = p(y,x)$. Also, this joint distribution is normalized:
\begin{equation}
\sum_{x \in \tilde{\mathcal{X}}} \sum_{y \in \tilde{\mathcal{Y}}}p(x,y) =1
\label{normqproduto}
\end{equation}

Acording to $GTS$ foundations, the entropy originally formulated by Tsallis is expressed as follows (\cite{Tsallis1988,Tsallis2009,GellMannTsallis}):

\begin{equation}
S_{q}(\tilde{\mathcal{X}}) = - \sum_{x \in \tilde{\mathcal{X}}} p^{q}(x) \ln_{q} p^{q}(x),
\label{Tentropia}
\end{equation}
where $p^{q}(x)$ is the generalized distribution associated with a discrete random variable $\tilde{\mathcal{X}}$. Yet, within the context of the $GTS$ the following deformed logarithm is introduced:
\begin{equation}
\ln_{q}(x)=\frac{x^{1-q}-1}{1-q}  \  \  \ (for \ \ x>0, q \in \mathbb{R})
\end{equation}

For systems A and B, Tsallis entropy is a non-additive entropy in general (\cite{Touchette2002,Tsallis2009}):
\begin{equation}
S_{q}^{(A+B)}=S_{q}^{(A)}+S_{q}^{(B)}+\frac{(1-q)}{k}S_{q}^{(A)}S_{q}^{(B)},
\end{equation}
where \textit{k} is the Boltzmann constant. In this sense $S_{q}$ is nonadditive for $q\neq 1$. Thus, we use expressions such as subadditive and superadditive, when $q > 1$ and $q < 1$ cases, respectively. In terms of discrete random variables $\tilde{\mathcal{X}}$ and $\tilde{\mathcal{Y}}$:
\begin{equation}
S_{q}(\tilde{\mathcal{X}},\tilde{\mathcal{Y}})=S_{q}(\tilde{\mathcal{X}})+S_{q}(\tilde{\mathcal{Y}})+\frac{(1-q)}{k}S_{q}(\tilde{\mathcal{X}})S_{q}(\tilde{\mathcal{Y}})
\end{equation}

Thus, we are able to define the joint Tsallis entropy, which gives a measure of a non-additive entropy of a composite system:

\begin{definition}
The joint $q$-entropy of a pair of discrete random variables $(\mathcal{\tilde{X},\tilde{Y}})$ with a joint distribution $p(x,y)$ is defined by:
\begin{equation}
S_{q}(\mathcal{\tilde{X},\tilde{Y}})= - \sum_{x \in \tilde{\mathcal{X}}} \sum_{y \in \tilde{\mathcal{Y}}} [p(x,y)]^{q} \ln_{q} p(x,y)
\label{joint_q_entropy}
\end{equation}
\end{definition}

While the joint Tsallis entropy gives a measure of the entropy of a composite system, the Shannon mutual information measures the degree of correlation between two systems, characterized by the probability distributions $p(x)$ and $p(y)$, being associated to random variables $\tilde{\mathcal{X}}$ and $\tilde{\mathcal{Y}}$, respectively. The Shannon mutual information, or simply the mutual information is defined as: 
\begin{definition}
The mutual information, $I(\mathcal{\tilde{X};\tilde{Y}})$ is the relative entropy between the joint distribution $p(x,y)$ and the product distribution $p(x)p(y)$ (\cite{Cover2006}):
\begin{equation}
I(\mathcal{\tilde{X};\tilde{Y}})=\sum_{x \in \tilde{\mathcal{X}}} \sum_{y \in \tilde{\mathcal{Y}}}p(x,y)  \ln \left[\frac{p(x,y)}{p(x)p(y)}\right]
\end{equation}
\end{definition}

We present a particular case of our formulation in which $p(x)$ and $p(y)$ are now the $q$-distributions $p_{q}(x)$ and $p_{q}(y)$, related to a pair of discrete random variables $(\mathcal{\tilde{X},\tilde{Y}})$. In this sense, we rewrite the Tsallis entropy, equation (\ref{Tentropia}) as: 
\begin{equation}
S_{q}(\tilde{\mathcal{X}}) = - \sum_{x \in \tilde{\mathcal{X}}} [p_{q}(x)]^{q} \ln_{q} p_{q}(x),
\end{equation}
where the $q$ is the same for the distributions $p_{q}(x)$, $p_{q}(y)$, the exponent and the $q$-logarithm. In the context of an interacting system, we also redefine the joint probability distribution, introducing the interaction index $q*$ in its formulation. Notice that $q*$ might differ from $q$, which generalizes our formulation as we now consider systems in which an entropic index related only to interactions might play a role. Therefore, the new joint probability is:
\begin{equation} 
p_{q*}(x,y)=\frac{p_{q}(x) \otimes_{q*} p_{q}(y)} {\sum_{y \in \tilde{\mathcal{Y}}}[p(x) \otimes_{q*} p(y)]}, \label{jointdist}
\end{equation}
Notice that the joint probability satisfy the redefined normalization condition:
\begin{equation}
\sum_{x \in \tilde{\mathcal{X}}} \sum_{y \in \tilde{\mathcal{Y}}}p_{q*}(x,y) =1 \label{sumjointdist0}
\end{equation}

The joint Tsallis entropy is also rewritten:
\begin{equation}
S_{q*}(\mathcal{\tilde{X},\tilde{Y}})= - \sum_{x \in \tilde{\mathcal{X}}} \sum_{y \in \tilde{\mathcal{Y}}} [p_{q*}(x,y)]^{q*} \ln_{q*} p_{q*}(x,y)
\end{equation}
where the $q*$ is the same for the joint distributions, the exponent and the $q$-logarithm.

And the Shannon mutual information is written as follows:
\begin{equation}
I_{qq*}(\mathcal{\tilde{X};\tilde{Y}})=\sum_{x \in \tilde{\mathcal{X}}} \sum_{y \in \tilde{\mathcal{Y}}}p_{q*}(x,y)  \ln \left[\frac{p_{q*}(x,y)}{p_{q}(x)p_{q}(y)}\right]
\end{equation}

From now on, we redirect the quantities and conditions defined previously to the study of ion channel kinetics. Suppose that the random variables $\tilde{\mathcal{X}}$ and $\tilde{\mathcal{Y}}$ are associated with the identical channels \textit{A} and \textit{B}. The channels present only two conformational states, closed $(C)$ and open $(O)$. Therefore, each random variable correspond to the same set of channel kinetic states, i.e. $\tilde{\mathcal{X}} = \tilde{\mathcal{Y}} = \{C, O\}$. Moreover, the probabilities of finding channels \textit{A} and \textit{B} in a given state are represented by the vector probabilities $p_{q}(x) = \textbf{p}_{q}^{A} = p_{q}(y) = \textbf{p}_{q}^{B} = ( p_{q,C} , p_{q,O} ) $

Using the presented notation, we rewrite the expressions of the Tsallis entropy as:
\begin{equation}
 S_{q} = S_{q}^{A} =  S_{q}^{B} = - \sum_{x \in {C,O}} [p_{q}(x)]^{q} \ln_{q} p_{q}(x)
                        = - (p_{q,C})^{q} \ln_{q} p_{q,C} - (p_{q,O})^{q} \ln_{q} p_{q,O}
\end{equation}

The joint distribution is now expressed as a vector:
\begin{eqnarray}
\textbf{p}_{qq*}^{AB} &=& \frac{1}{N} (\textbf{p}_{q}^{A} \otimes_{q*} \textbf{p}_{q}^{B}) \nonumber \\
                     &=& \frac{1}{N} [(p_{q,C} , p_{q,O}) \otimes_{q*} (p_{q,C} , p_{q,O})] \nonumber \\
                     &=& \frac{1}{N}[(p_{q,C} \otimes_{q*}  p_{q,C} \quad p_{q,C} \otimes_{q*}  p_{q,O} \quad
                          p_{q,O} \otimes_{q*}  p_{q,C} \quad p_{q,O} \otimes_{q*}  p_{q,O})] \nonumber \\
                     &=& (p_{qq*}^{CC} \quad  p_{qq*}^{CO} \quad p_{qq*}^{OC}  \quad p_{qq*}^{OO})                         
\label{jointdistvec}
\end{eqnarray}

Where \textit{N} is the normalization condition for the joint distribution:
\begin{equation}
N = p_{q,C} \otimes_{q*} p_{q,C} +  p_{q,C} \otimes_{q*} p_{q,O} + p_{q,O} \otimes_{q*} p_{q,C} + p_{q,O} \otimes_{q*} p_{q,O}  \label{normqprodutCO}
\end{equation}

The joint Tsallis entropy is rewritten as:
\begin{eqnarray}
S_{qq*}^{AB} &=& - \sum_{x \in \{C, O\}} \sum_{y \in \{C, O\}}[p_{qq*}(x,y)]^{q*} \ln_{q*} p_{qq*}(x,y) \nonumber \\
            &=& - (p_{qq*}^{CC})^{q*} \ln_{q*} p_{qq*}^{CC} - (p_{qq*}^{CO})^{q*}  \ln_{q*} p_{qq*}^{CO}
               - (p_{qq*}^{OC})^{q*}  \ln_{q*} p_{qq*}^{OC} - (p_{qq*}^{OO})^{q*}  \ln_{q*} p_{qq*}^{OO}
\label{qentropy}
\end{eqnarray}

Finally, we rewrite the mutual information as:
\begin{eqnarray}
I_{qq*}^{AB} &=& \sum_{x \in \{C, O\}} \sum_{y \in \{C, O\}}p_{q*}(x,y)  \ln \left[\frac{p_{q*}(x,y)}{p_{q}(x)p_{q}(y)}\right] \nonumber \\
            &=& p_{q*}^{CC} \ln \left[\frac{p_{q*}^{CC}}{p_{q,C}p_{q,C}}\right] 
              + p_{q*}^{CO} \ln \left[\frac{p_{q*}^{CO}}{p_{q,C}p_{q,O}}\right] \\ \nonumber 
            &+& p_{q*}^{OC} \ln \left[\frac{p_{q*}^{OC}}{p_{q,O}p_{q,C}}\right] 
              + p_{q*}^{OO} \ln \left[\frac{p_{q*}^{OO}}{p_{q,O}p_{q,O}}\right] 
\end{eqnarray}

As the joint probability is commutative, i.e $p_{qq*}^{C,O} = p_{qq*}^{O,C}$, it is possible to simplify the former expressions.

\subsection{Non-extensive ion channel probabilities}

Let us consider a pair of identical $MET$ channels, whose dynamics are treated as discrete-state Markov processes, in continuous-time, following non-extensive kinetics in the steady-state  (\cite{ColquhounHawkes1981}). The channels present only two conformational states, closed (\textit{C}) and open (\textit{O}). The probability of finding channel \textit{A} in a given state is represented by the vector probability $\textbf{p}_{q}^{A} = ( p_{q,C} , p_{q,O} ) $. 

The non-extensive steady-state probabilities of channels are dependent on $X$, the hair bundle's displacement from its resting position (in nm) (\cite{Hudspeth2000,Fettiplace2014}). These probabilities are inspired in work originally proposed by Erden in his single voltage channel analysis (\cite{Erdem2007}): 

\begin{equation}
p_{q,C}(X) = \frac{\exp_{q}\left(-z_{X}(X-X_{0})/k_{B}T\right)}{1 + \exp_{q}\left(-z_{X}(X-X_{0})/k_{B}T\right)}
\label{pclosedMET}
\end{equation}

and 

\begin{equation}
p_{q,O}(X) = \frac{1}{1 + \exp_{q}\left(-z_{X}(X-X_{0})/k_{B}T\right)},
\label{popenMET}
\end{equation}
where $z_{X}$ is the single-channel gating force (in pN), which measures mechanical sensitivity (\cite{Hudspeth2000,Fettiplace2014}). $X_{0}$ is  displacement for which the open probability is 50\% (in nm), $k_{B}$ is the Boltzmann constant and \textit{T} is the temperature (in Kelvin). 

The function $\exp_{q}$ in the former equations is the $q$-exponential function:
\begin{equation}
\exp_{q}(x) := \left[1+\left(1-q\right)x\right]^{1/(1-q)}, 
\end{equation}
if $1+\left(1-q\right) x\geq 0$ and $e_q^x=0$ otherwise. In the limit $q \rightarrow 1$ the usual exponential function is recovered. 

This formulation generalizes the usual expression for an ion channel steady-state kinetics, based on the Boltzmann-Gibbs statistics. It also introduces the $q$-exponential as a generalized function for the activation kinetics of $MET$ channels, linked to $q$. This assumption is crucial in our model for ion channel cooperation, which is based on a generalized joint probability distribution.

\begin{figure}
\centering
(a)\includegraphics[width=7.5cm]{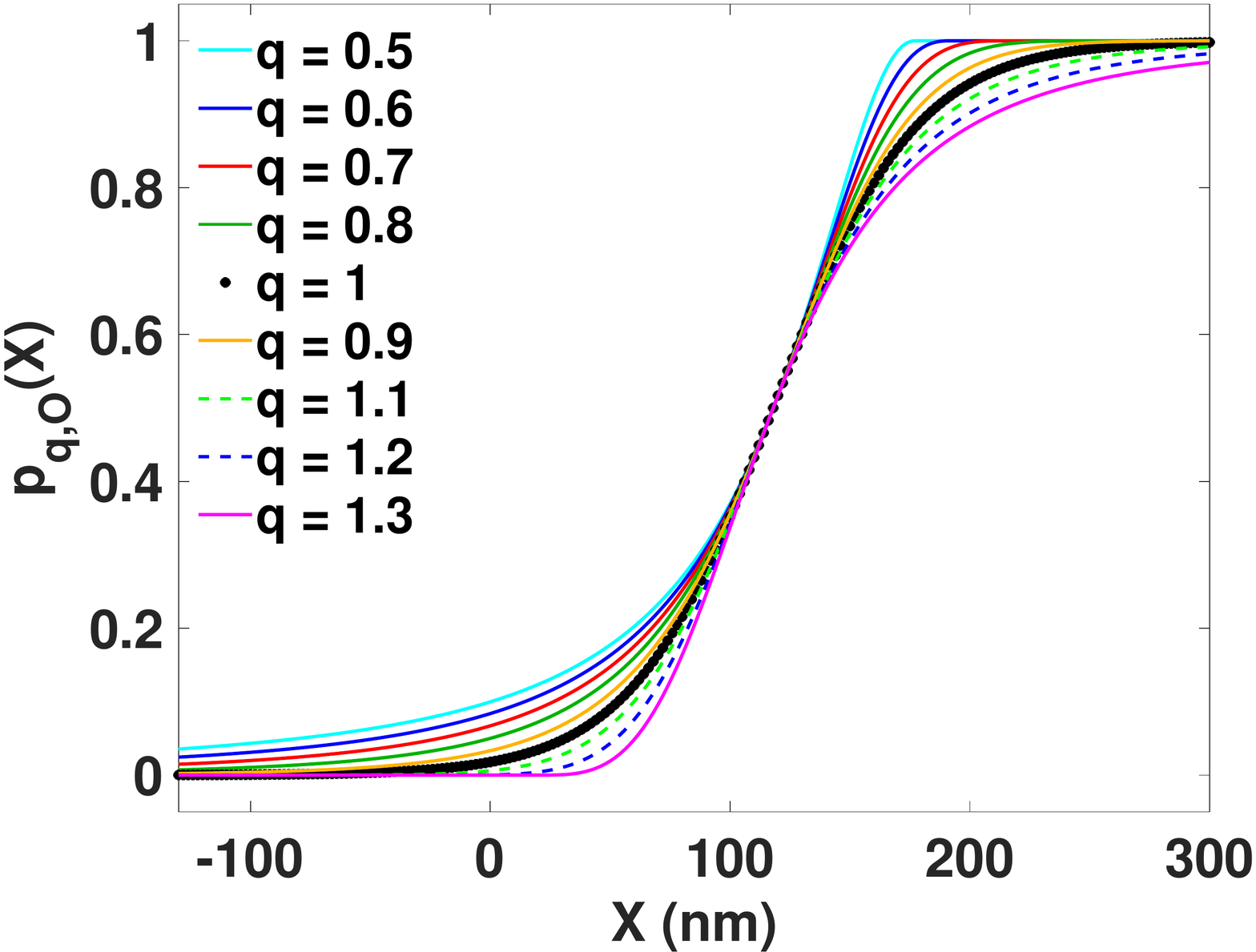}
(b)\includegraphics[width=7.5cm]{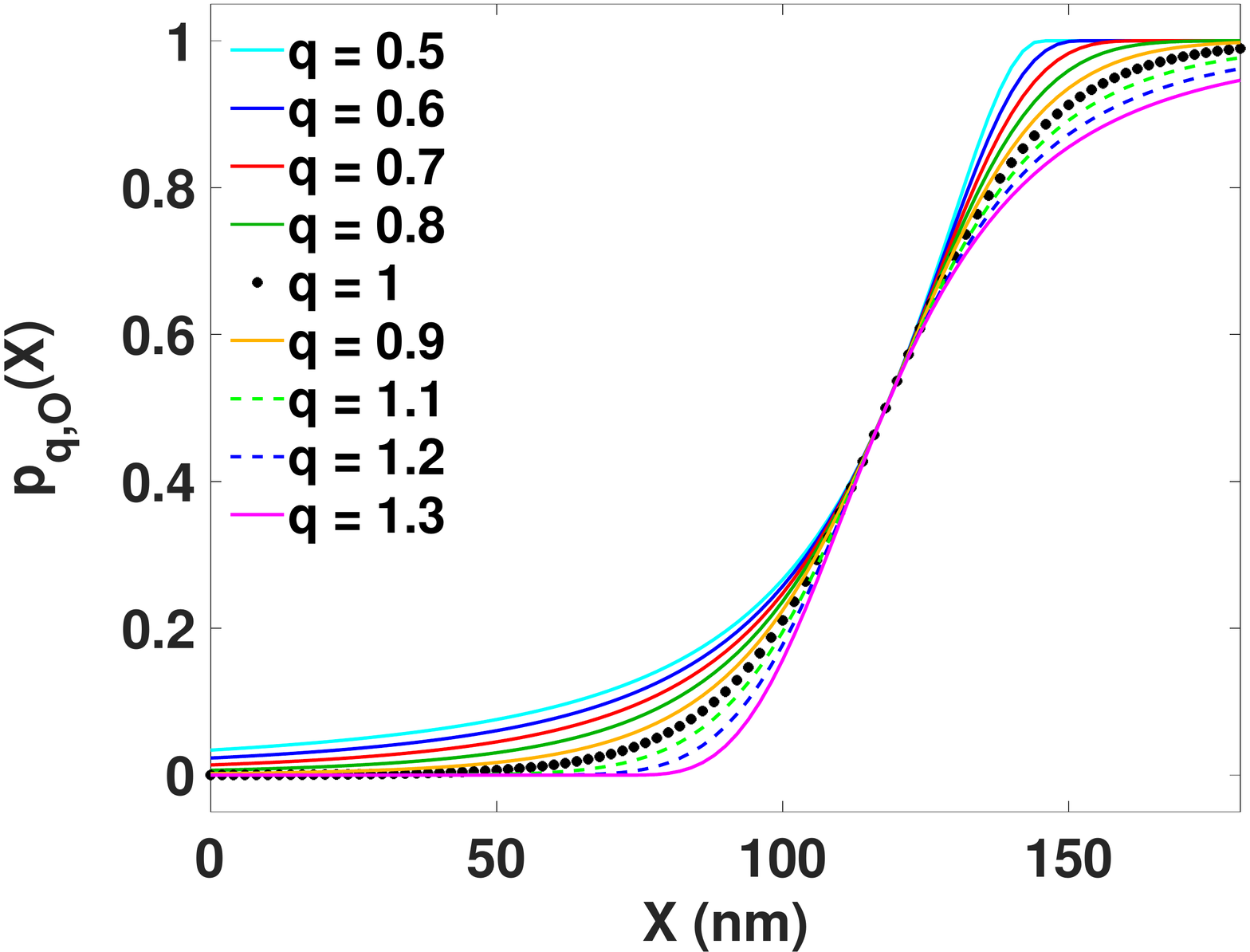}
\caption{Normalized open probability of a single nonextensive $MET$ channel, as a function of displacement, for several $q$ values.(a) $z_{X}$ = $0.138$ pN; (b) $z_{X}$ = $0.298$ pN. The $z_{X}$ parameter was extracted from (\cite{Hudspeth2000,Gianoli2017}). Displacement ranges were selected considering values used in experiments (\cite{Gianoli2017,Jia2007,He2004}). Other channel parameters are $X_{0}$ = $118$ nm, and \textit{T} = $21$°C extracted from (\cite{Howard1988,Hudspeth2000,Gianoli2017}).}
\label{figplotpopen}
\end{figure}

\section{Results}

\subsection{Activation probabilities and the Tsallis entropy}

We initially examined the activation probabilities for the non-extensive channels, searching the individual $q$ values for which the curves have a sigmoidal shape (figure \ref{figplotpopen}). For comparison, the extensive open probability ($q$ = 1.0) is shown in black dot style. For the unitary gating forces $z_{X}$ = $0.138$ pN and $0.298$ pN, the $q$ belongs to the interval [$0.5$-$1.3$], if we keep the remaining parameters, as the activation half displacement ($X_{0}$ and temperature $(T)$ fixed. Changing $X_{0}$ led to a displacement of the activation curves, while modification of the temperature in the interval [$21.0$°C-$37.0$°C] minimally altered the shape of the curves. Further modification of $z_{X}$ or any other parameters did not influence the activation curve shapes for a given $q$ value. For $q < 1.0$ (solid lines), the activation curves appear above the extensive activation curve in black, which means that the channels present a higher open probability for lower $X$ values. The curves associated with $q > 1.0$ (dashed lines) tend to present lower open probabilities for higher $X$, but the curves present slight deviations from the extensive curve. Moreover, all curves cross at $X_{0}$ = 118 nm, where the activation probability is 50$\%$.
\begin{figure}
\centering
(a)\includegraphics[width=7.5cm]{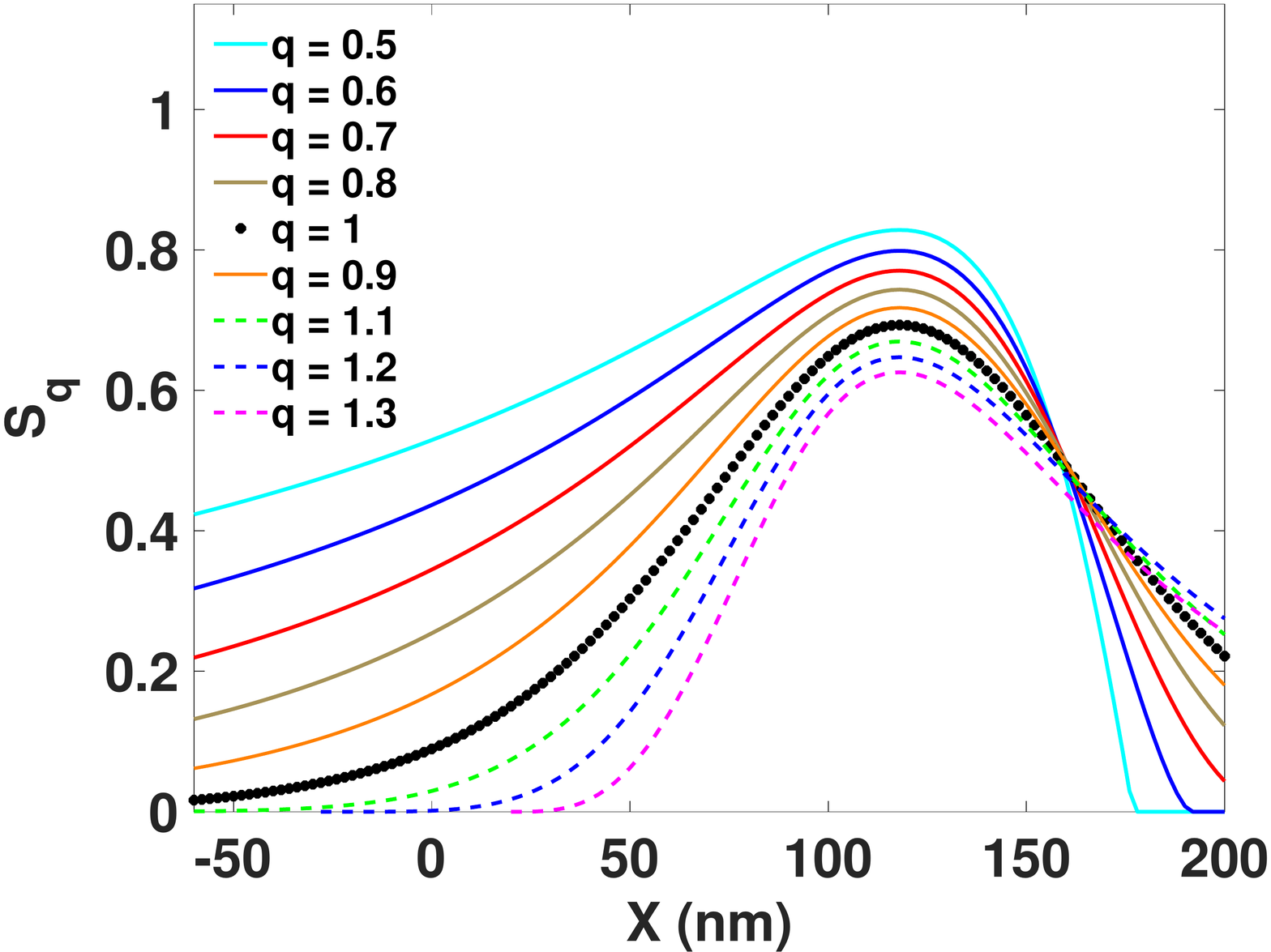}
(b)\includegraphics[width=7.5cm]{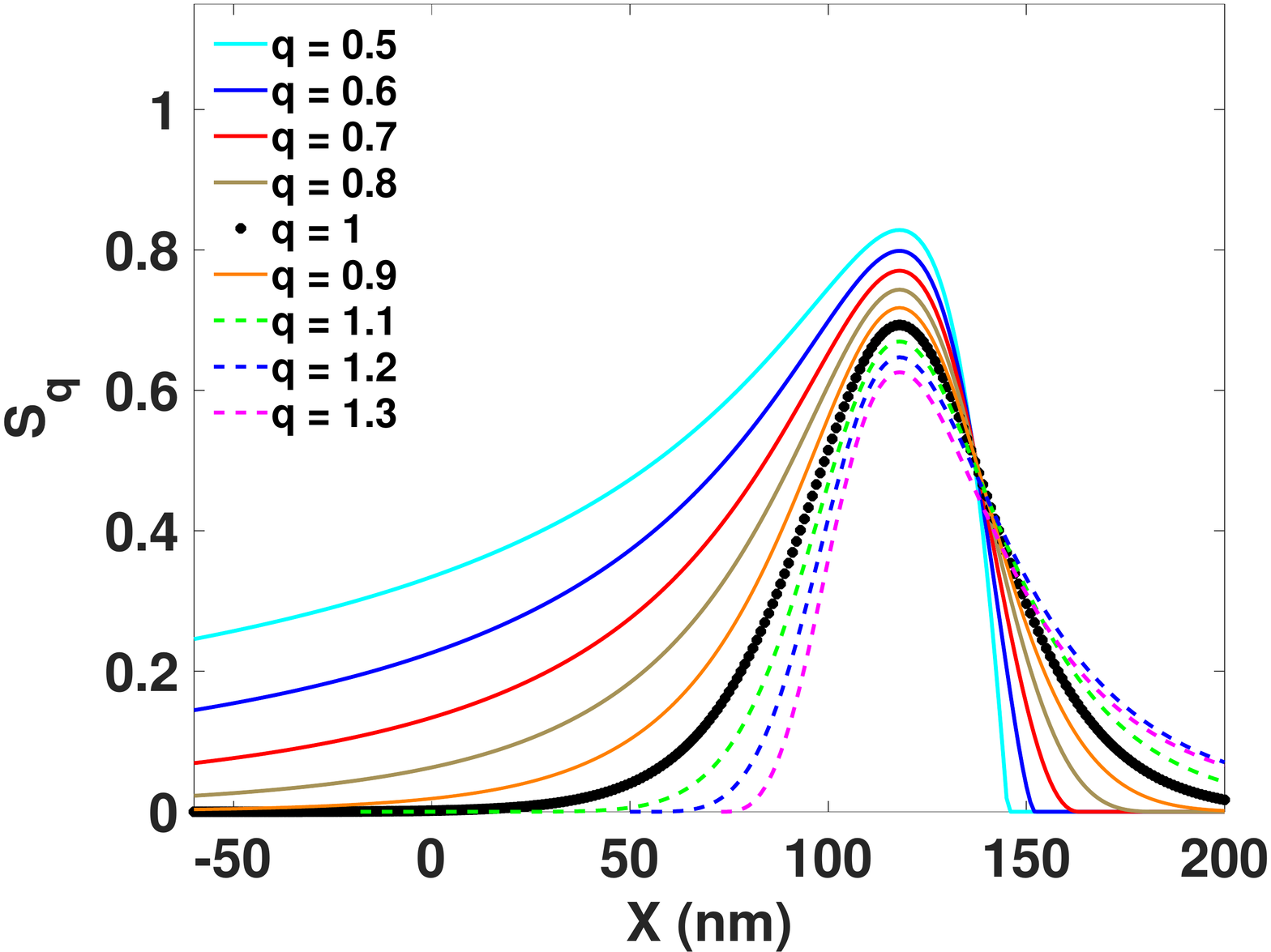}
\caption{Tsallis entropy as a function  of displacement for several $q$ values. Other channel parameters are $z_{X}$ = $0.138$ pN (a), $z_{X}$ = $0.298$ pN (b), $X_{0}$ = $118$ nm, and \textit{T} = $21$°C extracted from (\cite{Howard1988,Hudspeth2000,Gianoli2017}).}
\label{figSq}
\end{figure}

In figure \ref{figSq} we calculated the single ion channel Tsallis entropy ($S_q$) assuming various individual $q$ values under and above the extensive case $q=1$. Comparing this figure with the plots for the activation probability (figure \ref{figplotpopen}), we observe that $q<1$ imply in higher values for $S_q$ than in the extensive case and lower values if $q>1$. Also, for $q<1$, the curve's increase is slow, but the curve presents a steeper rise if one takes $q>1$. All peaks are located at the $X_{0}$ value $118$ nm, in which the normalized open probability has a value of $0.5$ (where open and close probabilities are equal). This happens for all values of the individual $q$. The curves tend to cross in the interval $158 < X < 162$ nm, for $z_{X}$ = $0.138$ pN, and $135 < X < 140$ nm for $z_{X}$ = $0.298$ pN, where the open probability is between 80$\%$ and 90$\%$ for $q > 1$ and above 90$\%$ for $q \geq 1$. After this interval, $S_q$ amplitudes are lower than the extensive curve ($q=1$) and decay rapidly for $q < 1$. For $q > 1$, $S_q$ values are higher than the extensive curve and decay more slowly, the slower the decay for increasing $q$. Still, according to figure \ref{figSq}, we can observe, for the values situated to the left of $X_{0}$, that the entropy will be greater for the smaller $z_{X}$ value. At the same time, it remains unchanged in relation to the values situated to the right of $X_{0}$. Surprisingly, these results suggest a connection between $z_{X}$ and entropic degree. 

\subsection{Joint probabilities, entropies and mutual information}

\begin{figure}
\begin{subfigure}{.32\textwidth}
\centering
(a) \includegraphics[width=1.0\linewidth]{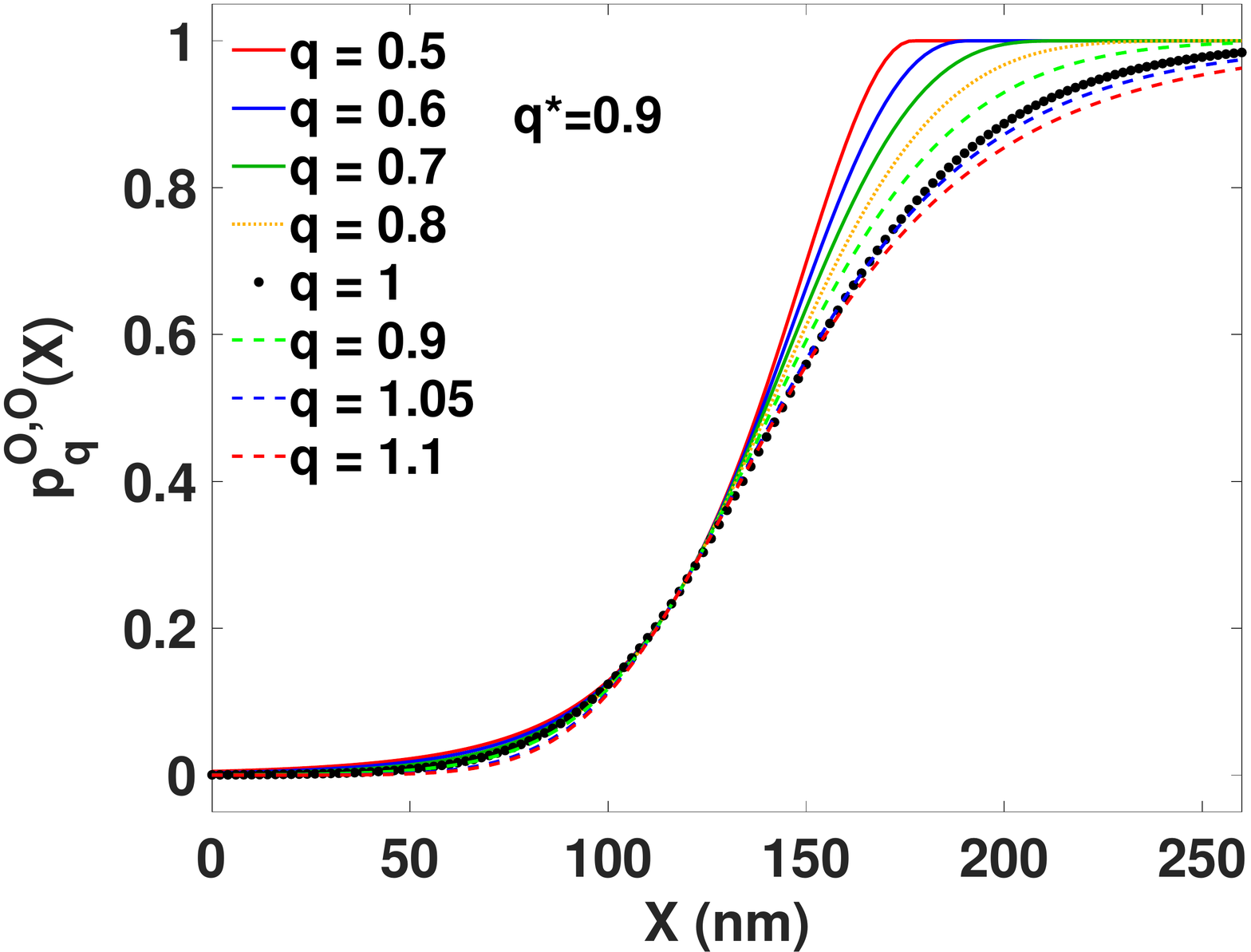} 

\end{subfigure}
\begin{subfigure}{.32\textwidth}
\centering
(b)  \includegraphics[width=1.0\linewidth]{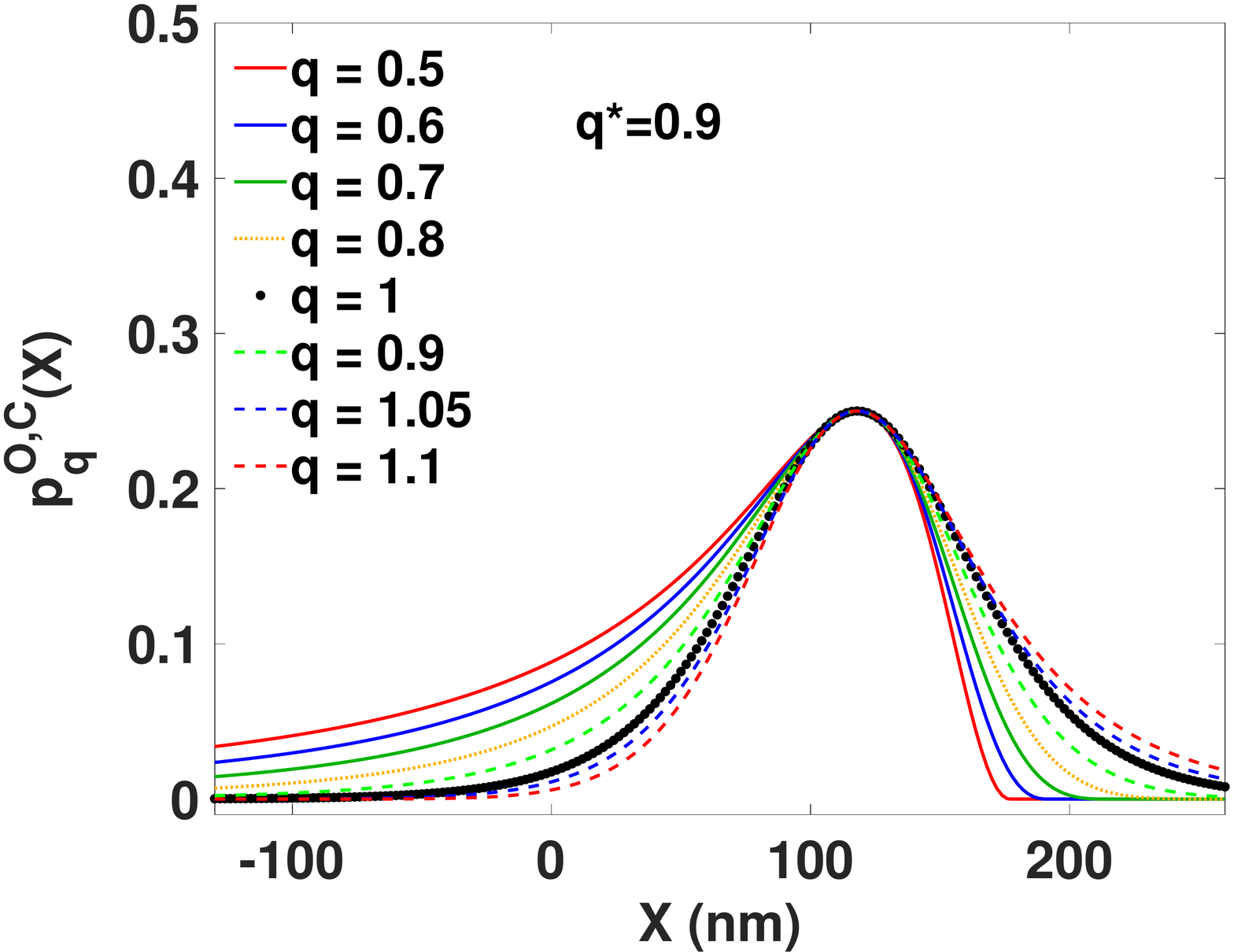}  
\end{subfigure}
\begin{subfigure}{.32\textwidth}
\centering
(c)  \includegraphics[width=1.0\linewidth]{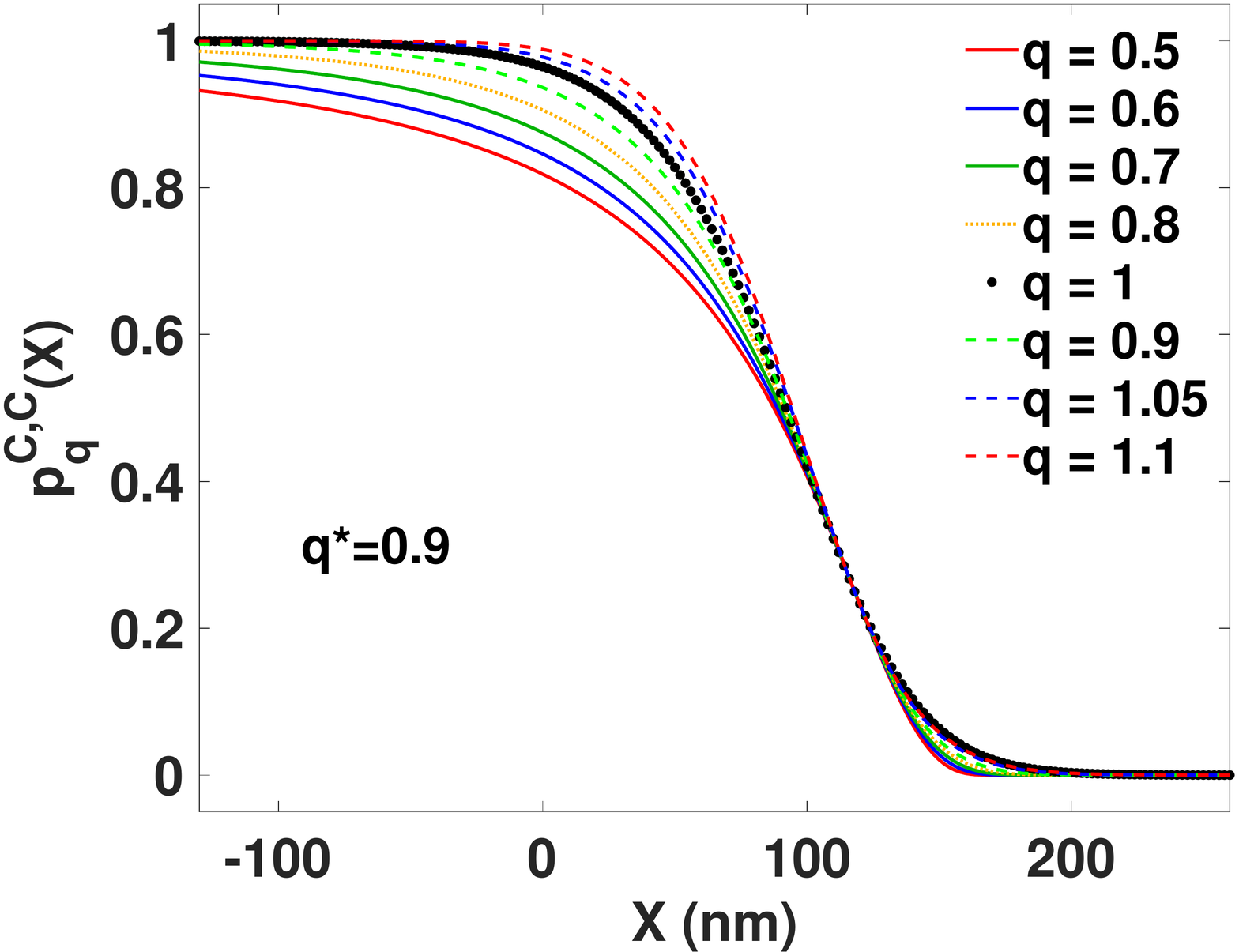}  
\end{subfigure}

\begin{subfigure}{.32\textwidth}
\centering
\vspace{0.2cm}
(d)  \includegraphics[width=1.0\linewidth]{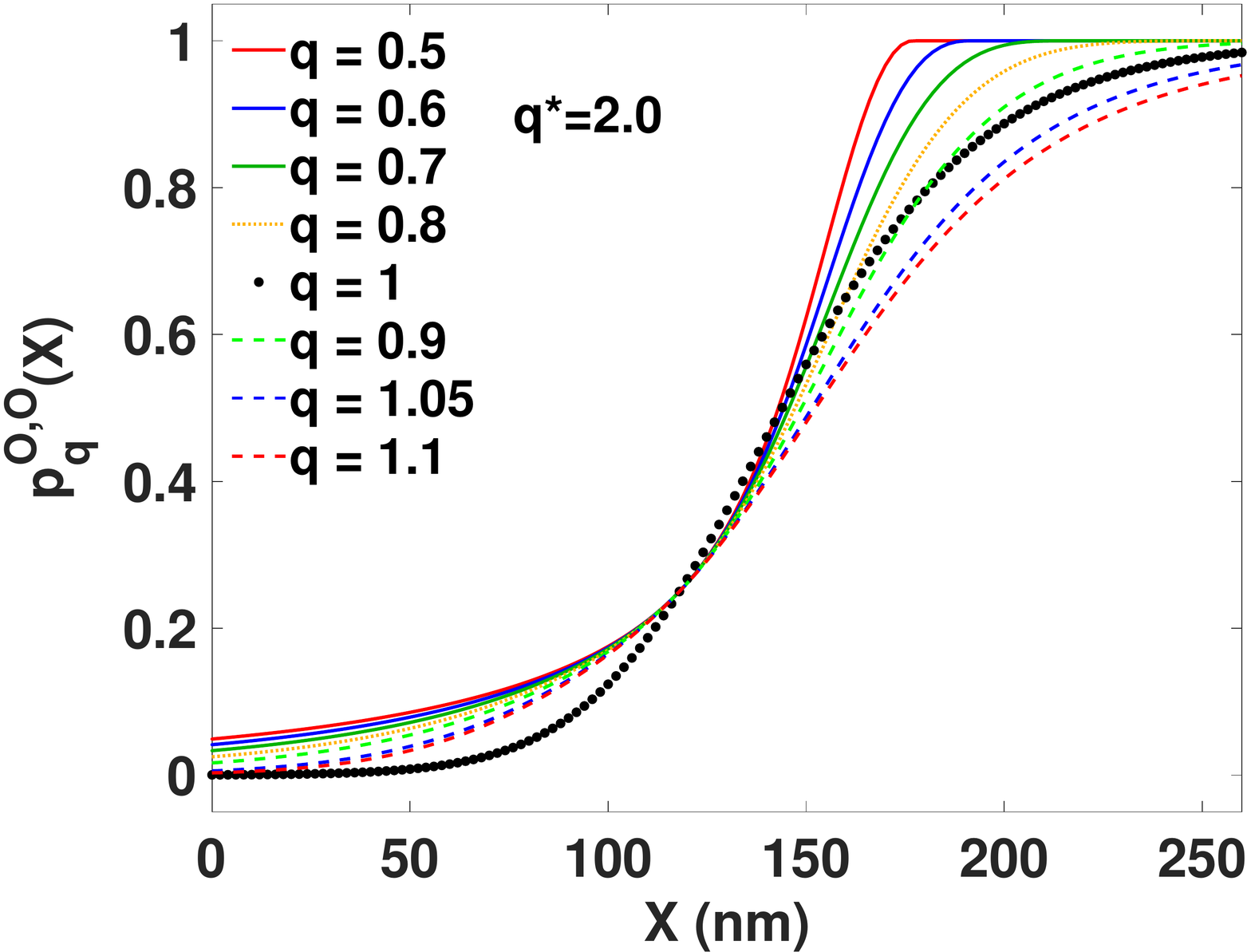}  
\end{subfigure}
\begin{subfigure}{.32\textwidth}
\centering
\vspace{0.2cm}
(e)  \includegraphics[width=1.0\linewidth]{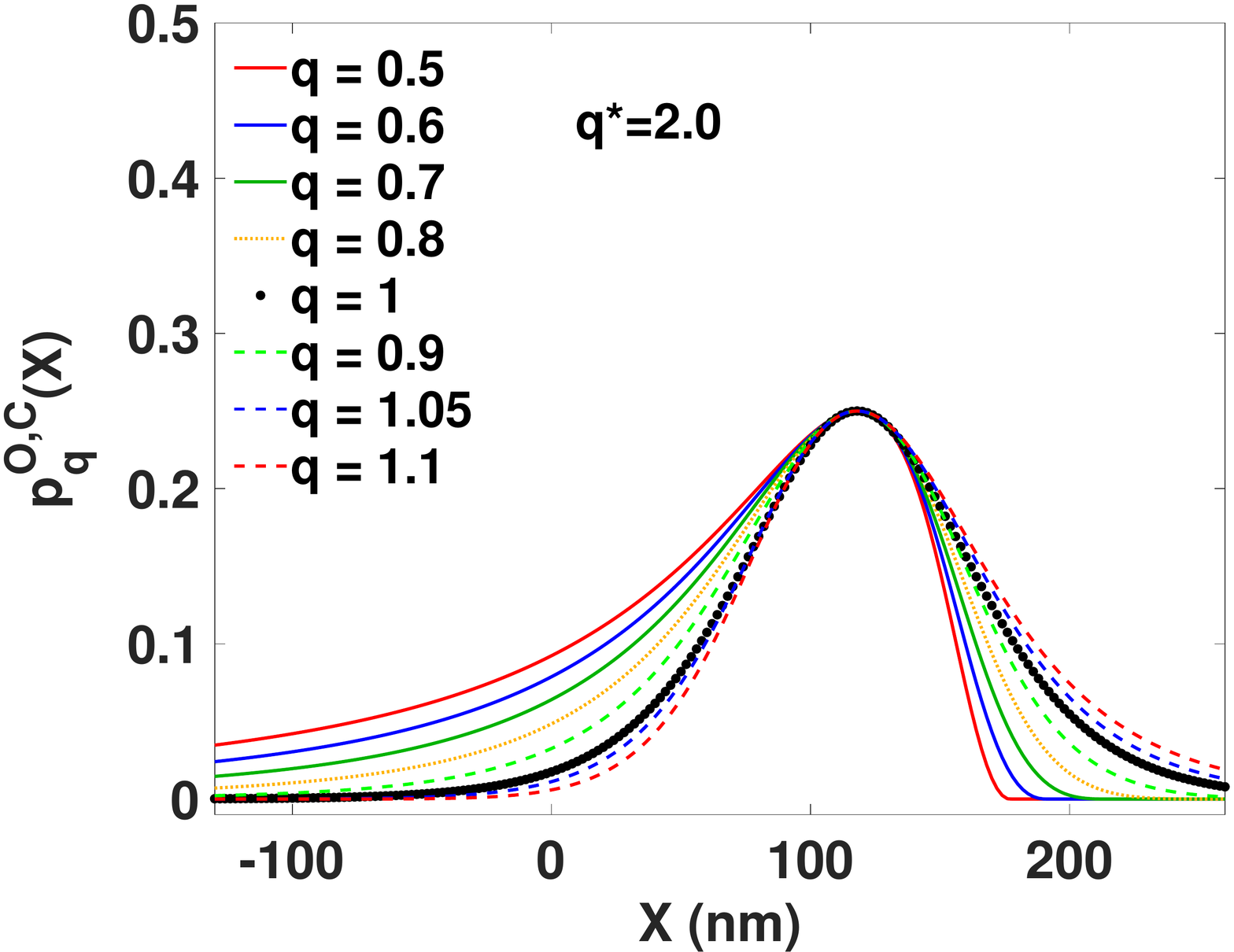}  
\end{subfigure}
\begin{subfigure}{.32\textwidth}
\centering
\vspace{0.2cm}
(f)  \includegraphics[width=1.0\linewidth]{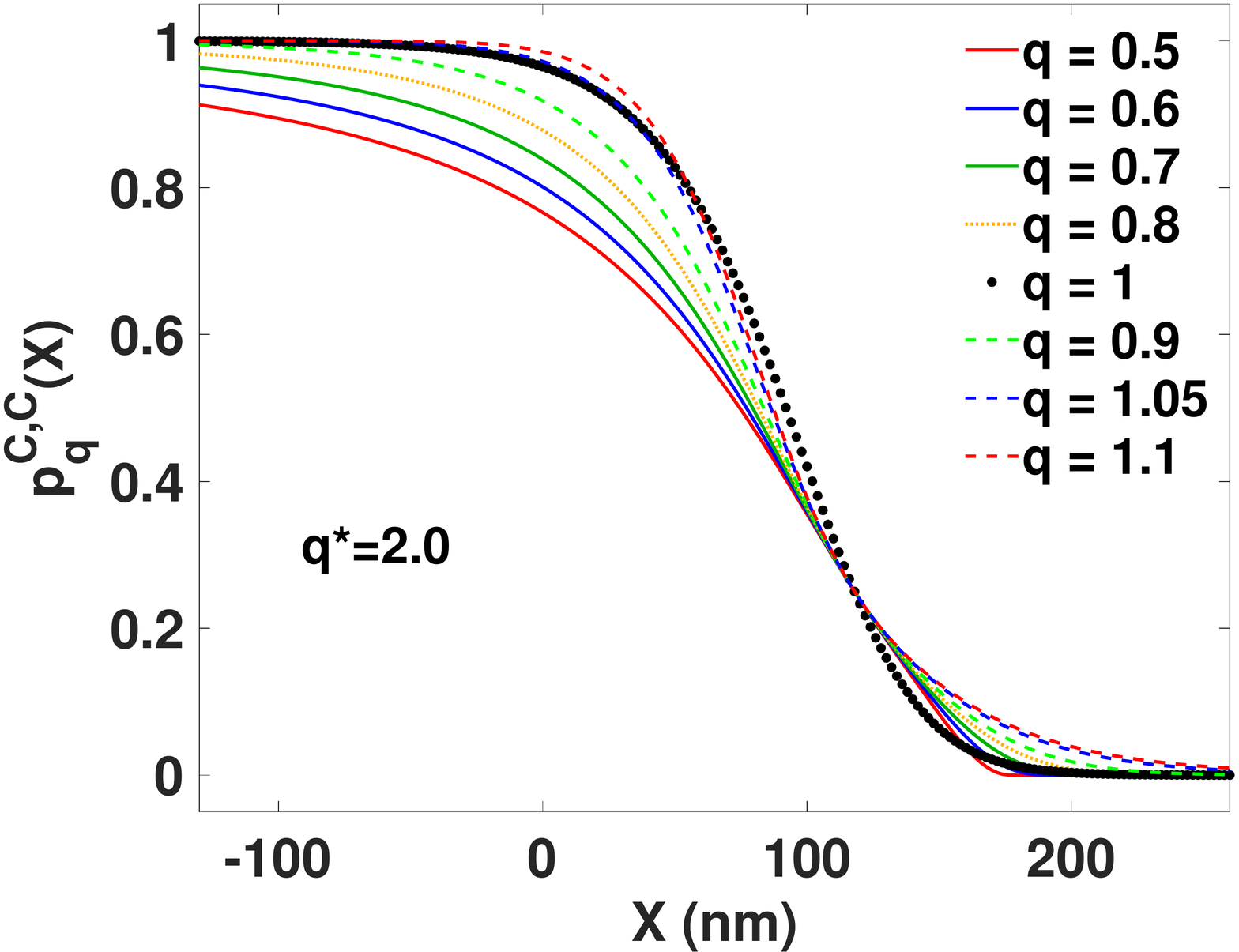}  
\end{subfigure}

\begin{subfigure}{.32\textwidth}
\centering
\vspace{0.2cm}
(g)  \includegraphics[width=1.0\linewidth]{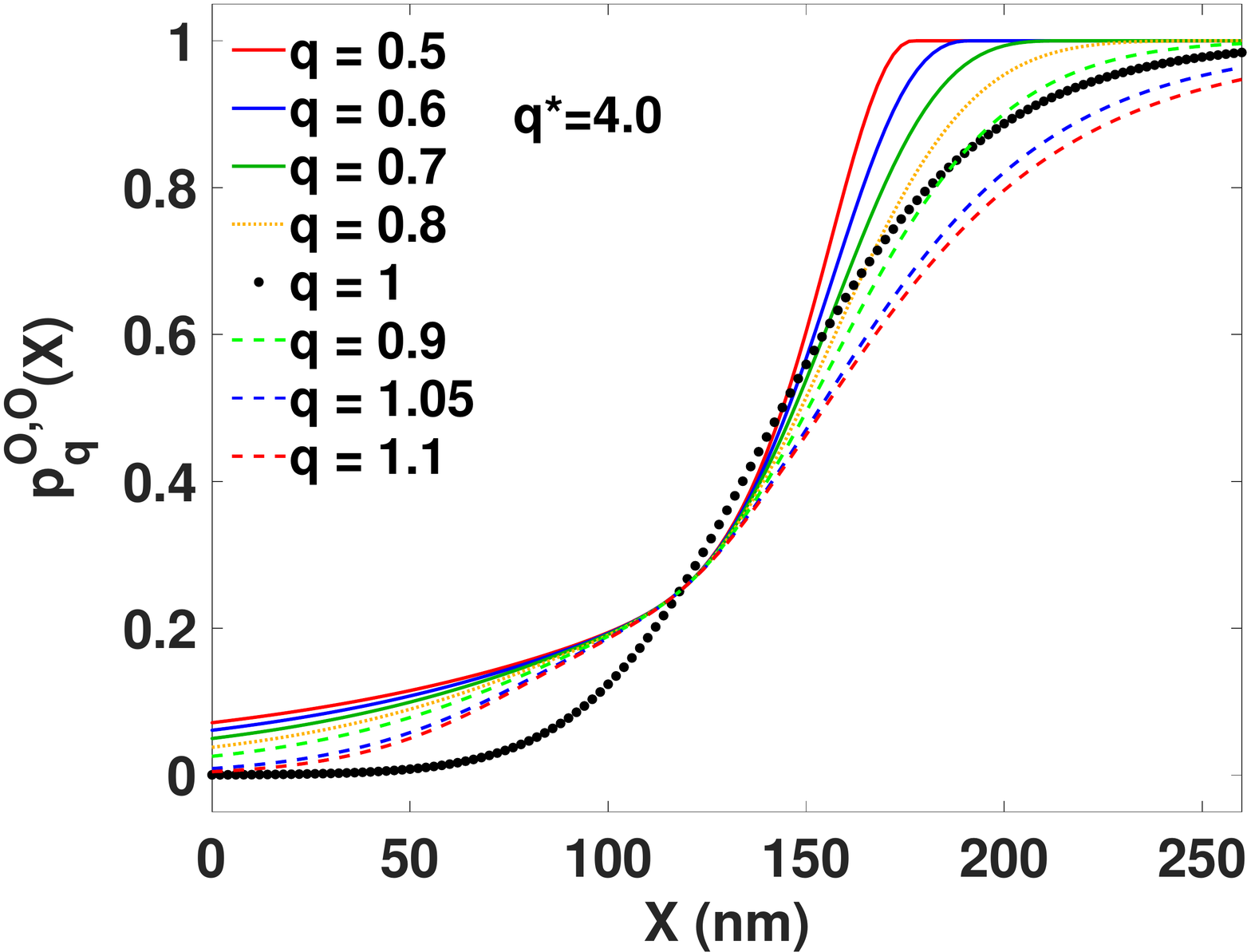}  
\end{subfigure}
\begin{subfigure}{.32\textwidth}
\centering
\vspace{0.2cm}
(h)  \includegraphics[width=1.0\linewidth]{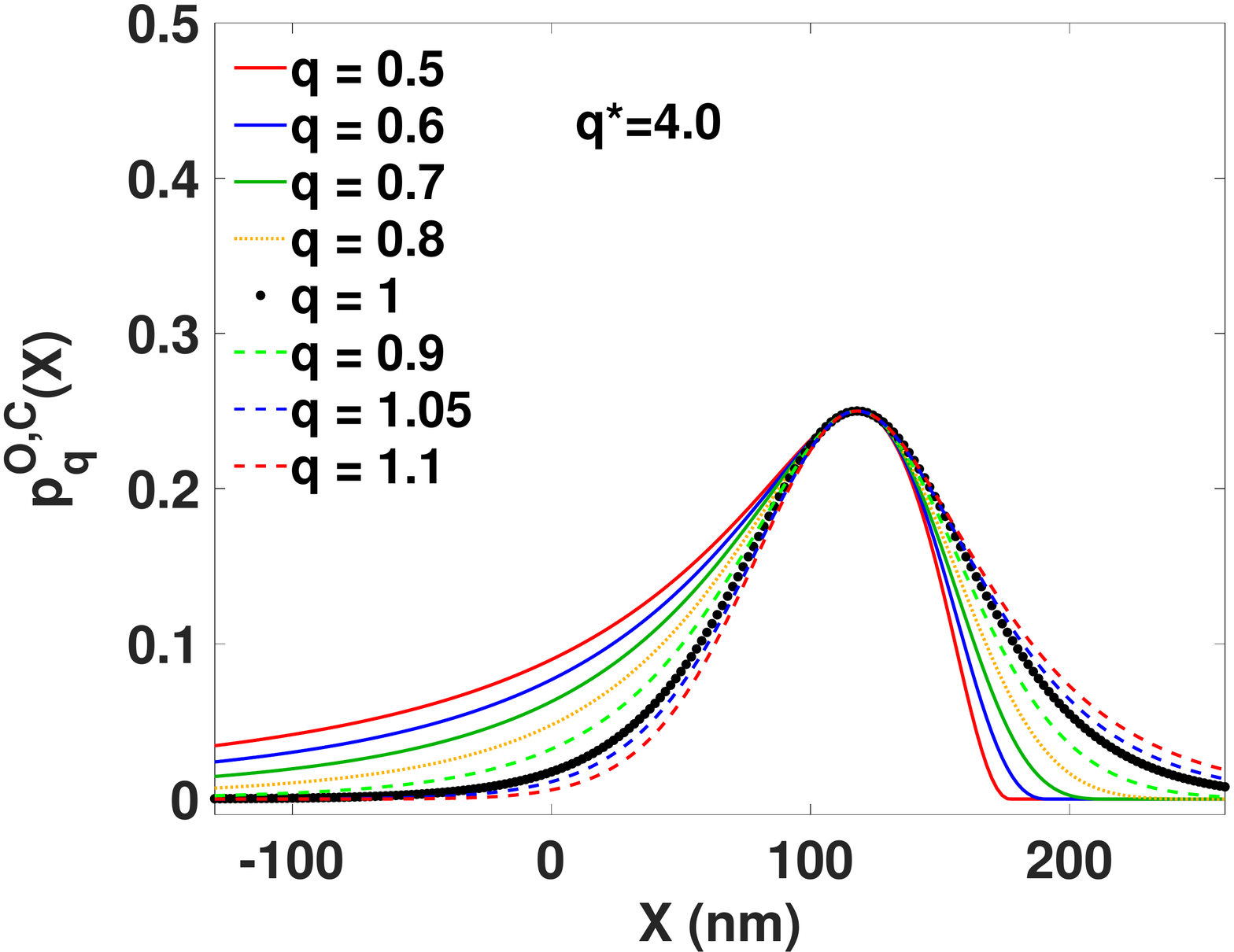}  
\end{subfigure}
\begin{subfigure}{.32\textwidth}
\centering
\vspace{0.2cm}
(i)  \includegraphics[width=1.0\linewidth]{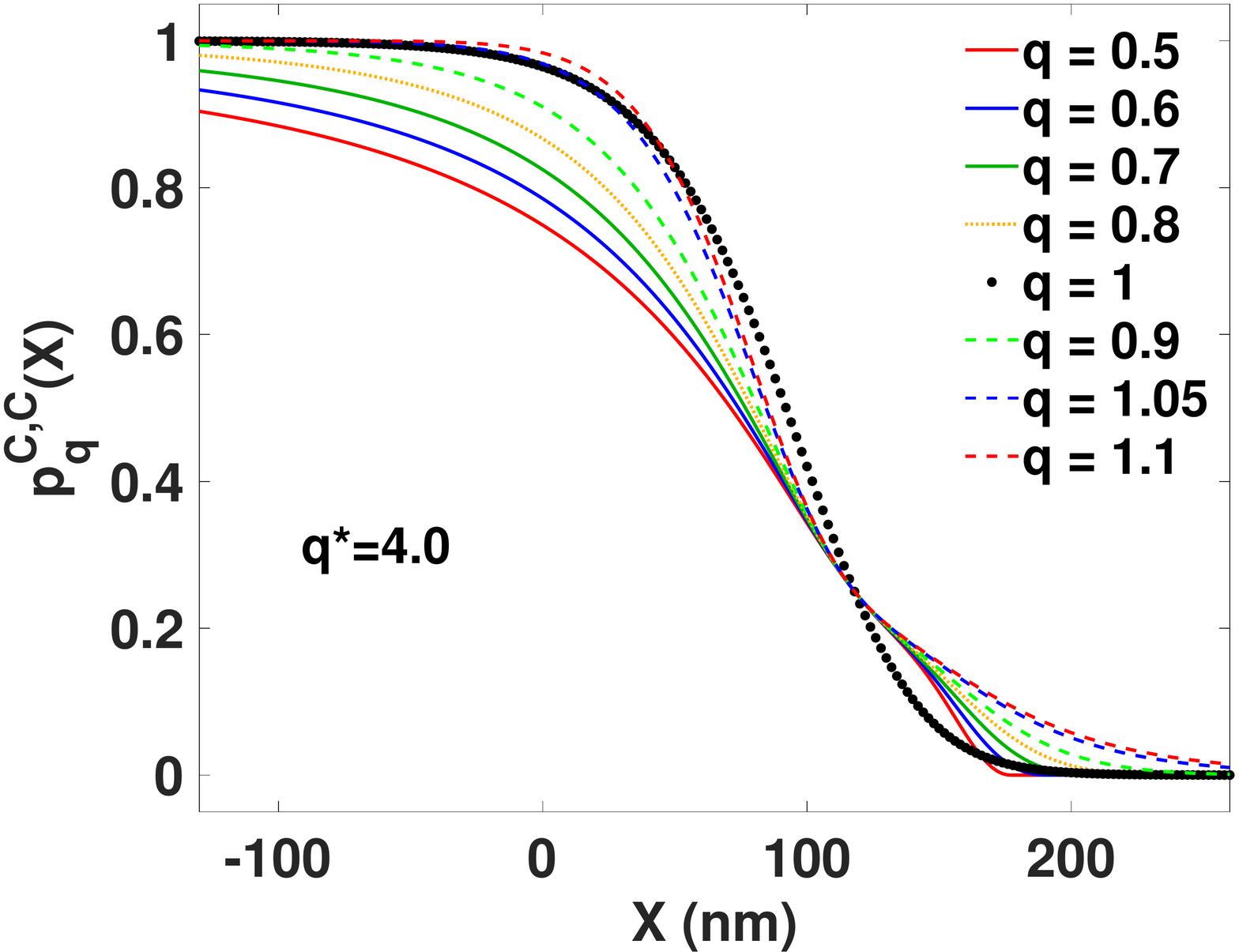}  
\end{subfigure}

\centering
\caption{Joint probabilities for a pair of interacting and identical $MET$ channels, as functions of the displacement for $qs = 0.9$ (above), $q* = 2.0$ (middle), $q* = 
4.0$ (below) for several $q$ values. Left: joint probability of open probability distributions, $p_{q}^{O,O}$  (plots (a), (d), (g)). Center: joint probability of open and closed probability distributions, $p_{q}^{O,C}$  (plots (b), (e), (h)). Right: joint probability of closed probability distributions $p_{q}^{C,C}$ (plots (c), (f), (i)). Other channel parameters are $z_{X}$ = $0.138$ pN, $X_{0}$ = $118$ nm, and \textit{T} = $21$°C extracted from (\cite{Howard1988,Hudspeth2000,Gianoli2017}).
}
\label{figplotqprodut}
\end{figure}

We next built plots of the joint probability distributions as functions of the displacement $(X)$, for various individual $q$ and $q*$. We did the individual $q$ the same for both channels. In figure \ref{figplotqprodut} the plots represent the joint probability distributions for the pair of channel states: open and open ($p^{oo}_{q}(X)$ - plots (a),(d),(g)), open and closed ($p^{oc}_{q}(X)$ -plots(b),(e),(h)), and closed and open ($p^{cc}_{q}(X)$ -plots (c),(f),(i)). We observe a sigmoid form for both activation,  $p^{oo}_{q}(X)$ and deactivation, $p^{cc}_{q}(X)$ curves in the range $0.5 < q < 1.1$. In contrast, the curve for open and closed probability, $p^{oc}_{q}(X)$, has a bell shape, if we fix the remaining parameters. Both activation and deactivation curves, for $q$ values cross at $X = X_{0}$, displacement at which the open and closed probabilities have a peak.

The activation curves present increasing values for decreasing $q$ for all $X$ and $q*$. For $X < X_{0}$ and $q* > 1$ we note that the non-extensive open probabilities are greater than the extensive ones. The open and closed probability plots present the same behavior, regardless of the interaction $q*$: for $X < X_{0}$ and decreasing $q$ corresponds to increasing $p^{oc}_{q}$. On the other hand, for $X > X_{0}$ this behavior inverts, and $q$ decrement is associated to decaying of $p^{oc}_{q}$ and a steeper falling for $q < 1$.

\begin{figure}
\begin{subfigure}{.32\textwidth}
\centering
(a) \includegraphics[width=0.9\linewidth]{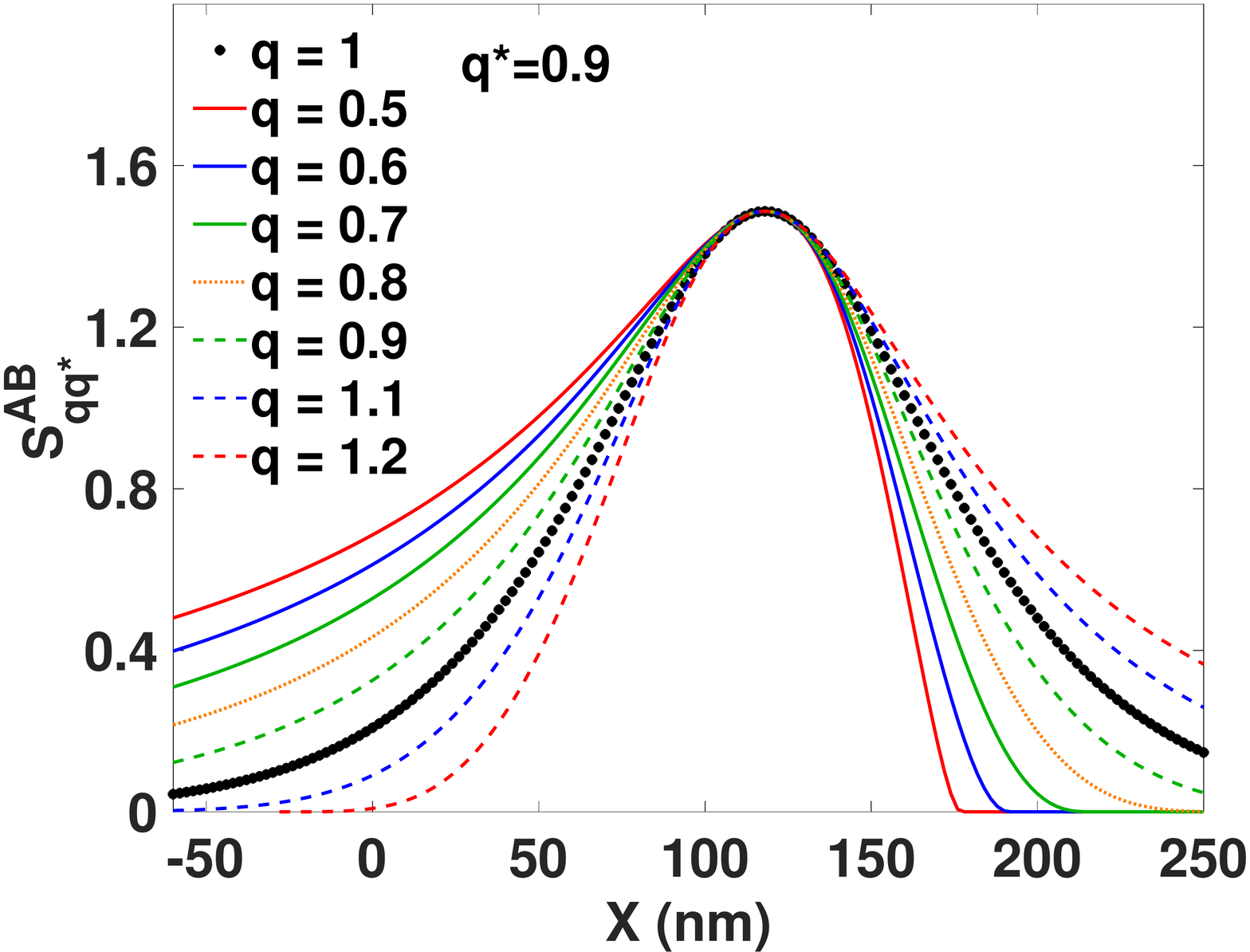}  
\end{subfigure}
\begin{subfigure}{.32\textwidth}
\centering
(b)\includegraphics[width=0.9\linewidth]{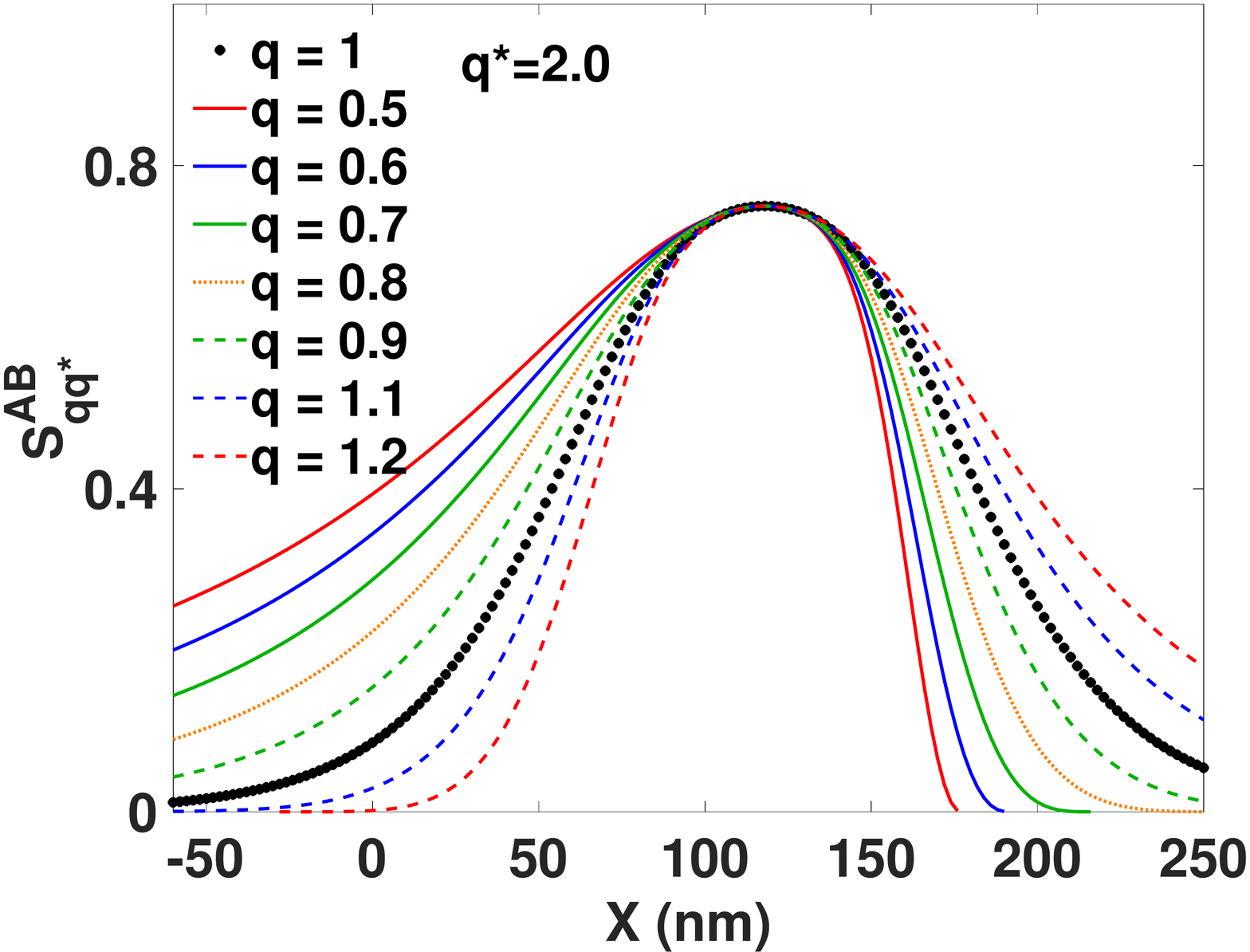}  
\end{subfigure}
\begin{subfigure}{.32\textwidth}
\centering
(c)  \includegraphics[width=0.9\linewidth]{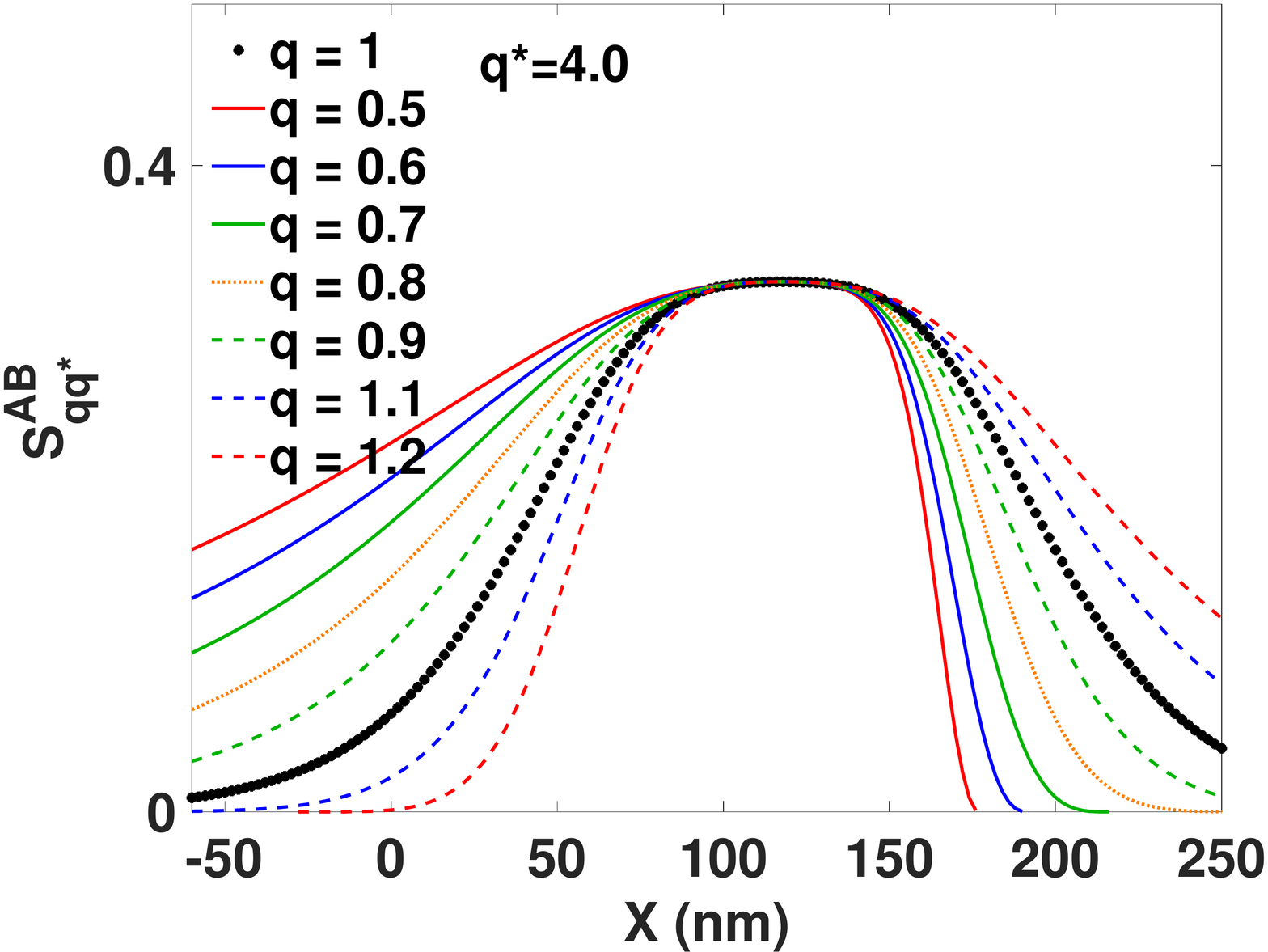}  
\end{subfigure}

\begin{subfigure}{.32\textwidth}
\centering
\vspace{0.2cm}
(d)  \includegraphics[width=0.9\linewidth]{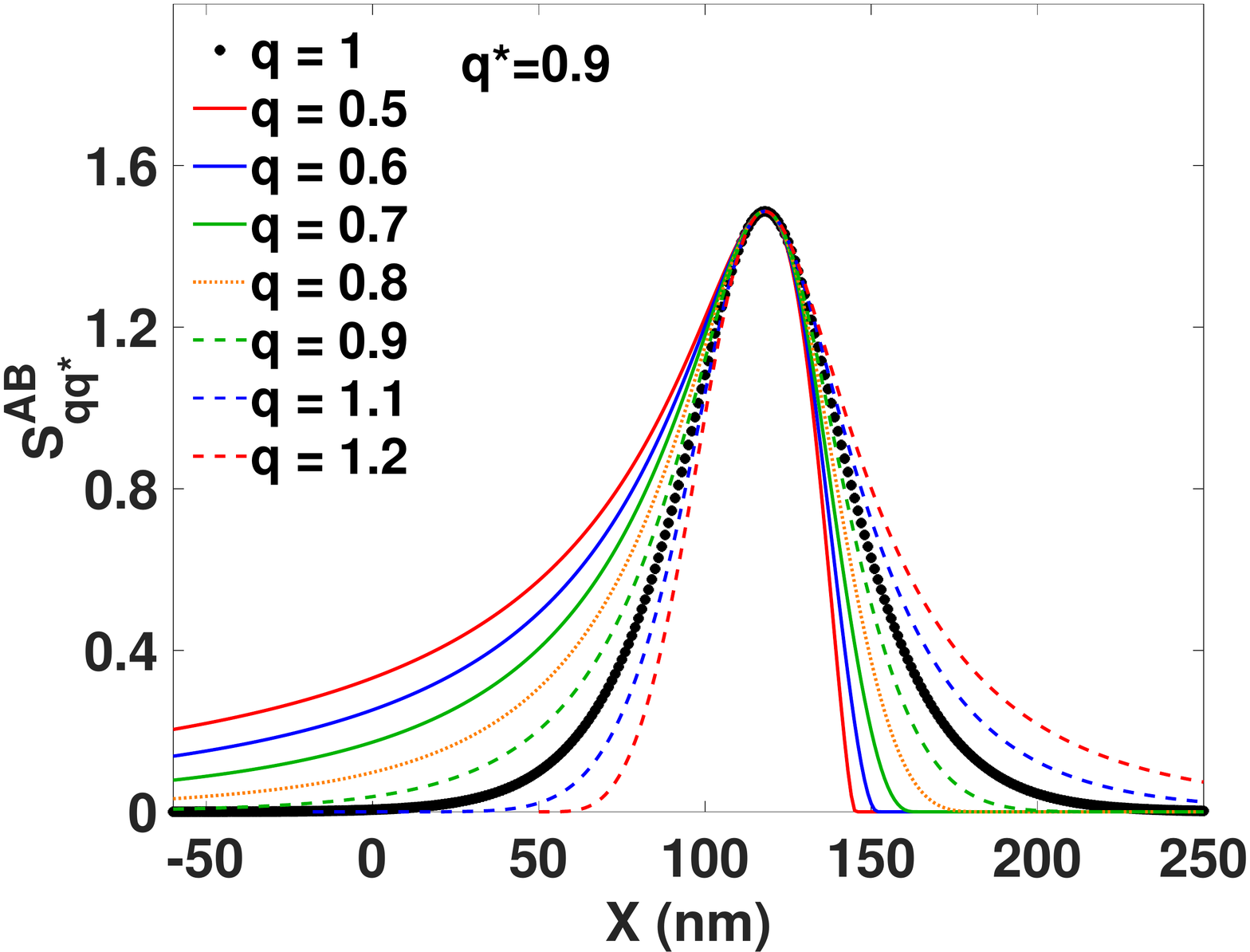}  
\end{subfigure}
\begin{subfigure}{.32\textwidth}
\centering
\vspace{0.2cm}
(e)  \includegraphics[width=0.9\linewidth]{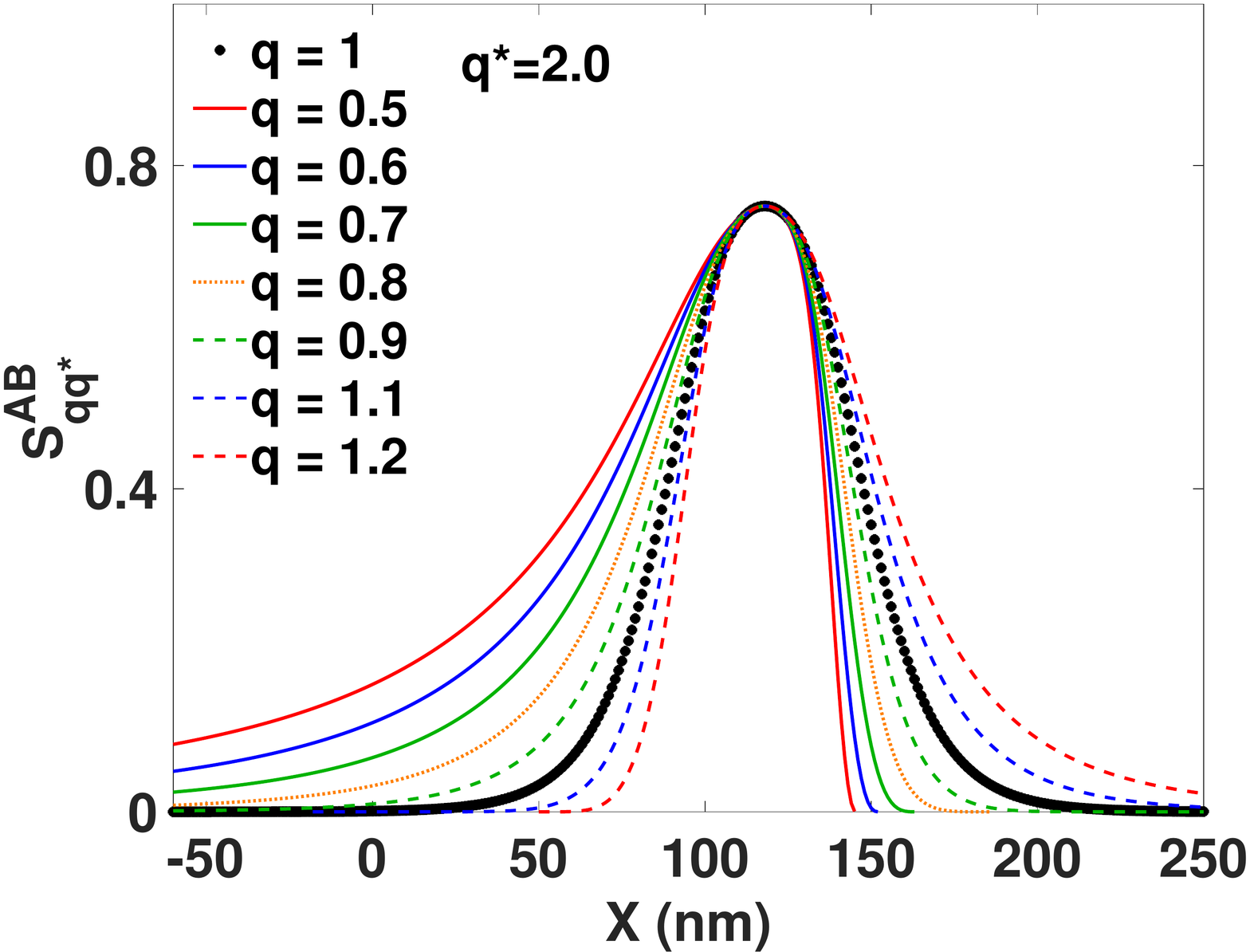}  
\end{subfigure}
\begin{subfigure}{.32\textwidth}
\centering
\vspace{0.2cm}
(f)  \includegraphics[width=0.9\linewidth]{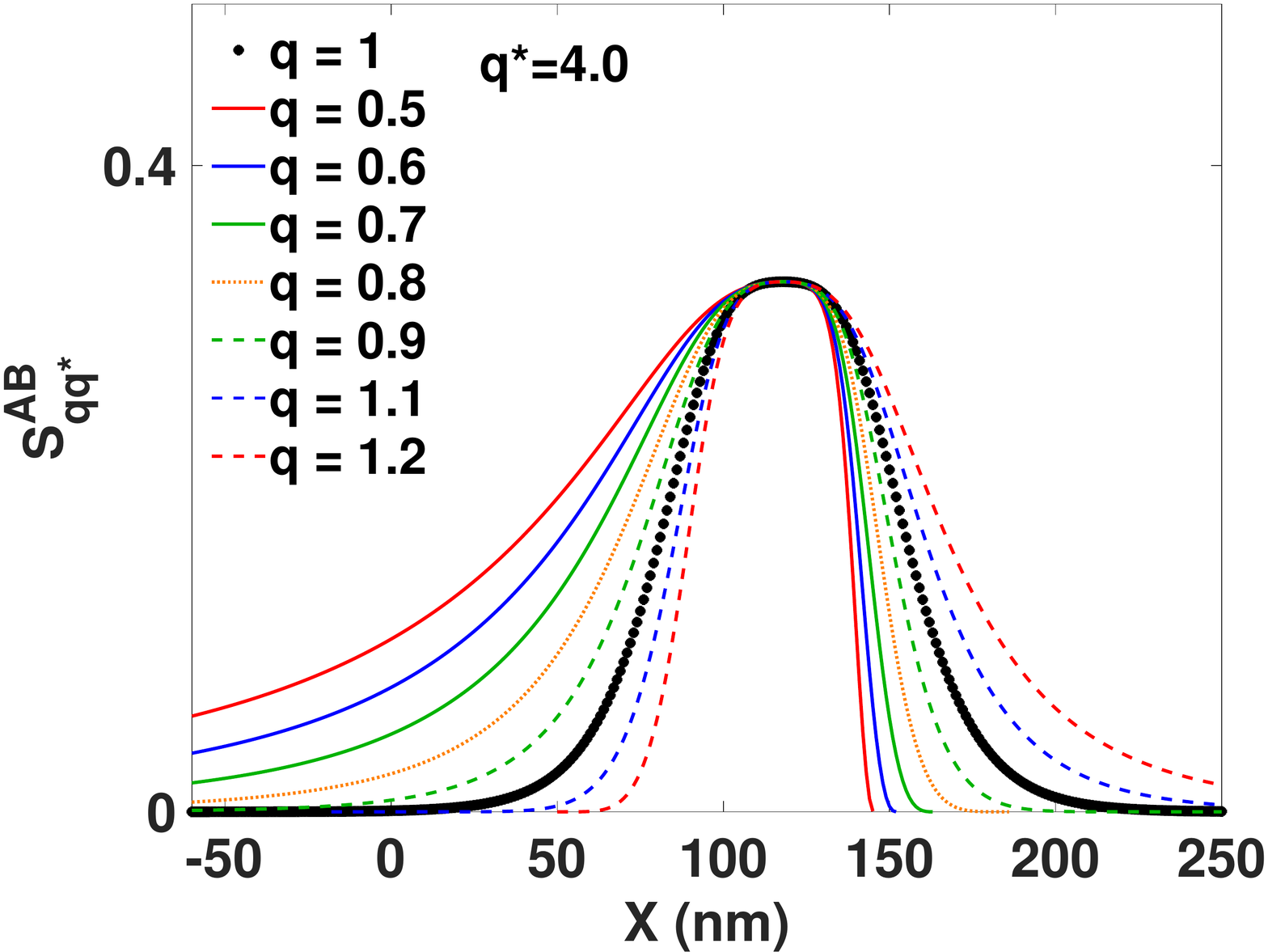}  
\end{subfigure}

\centering
\vspace{0.2cm}
\caption{Tsallis joint entropies for a pair of interacting and identical $MET$ channels, as functions of the displacement for $q* = 0.9$ (a)-(d), $q* = 2.0$ (b)-(e), $q* = 4.0$ (c)-(f) and several $q$ values. Other channel parameters are: $z_{X}$ = $0.138$ pN is used in (a)-(b)-(c), $z_{X}$ = $0.298$ pN is assumed in (e)-(f)-(g), $X_{0}$ = $118$ nm, and \textit{T} = $21$°C extracted from (\cite{Howard1988,Hudspeth2000,Gianoli2017}).}
\label{figSqconjunta}
\end{figure}

\begin{figure}
\begin{subfigure}{.32\textwidth}
\centering
(a) \includegraphics[width=0.9\linewidth]{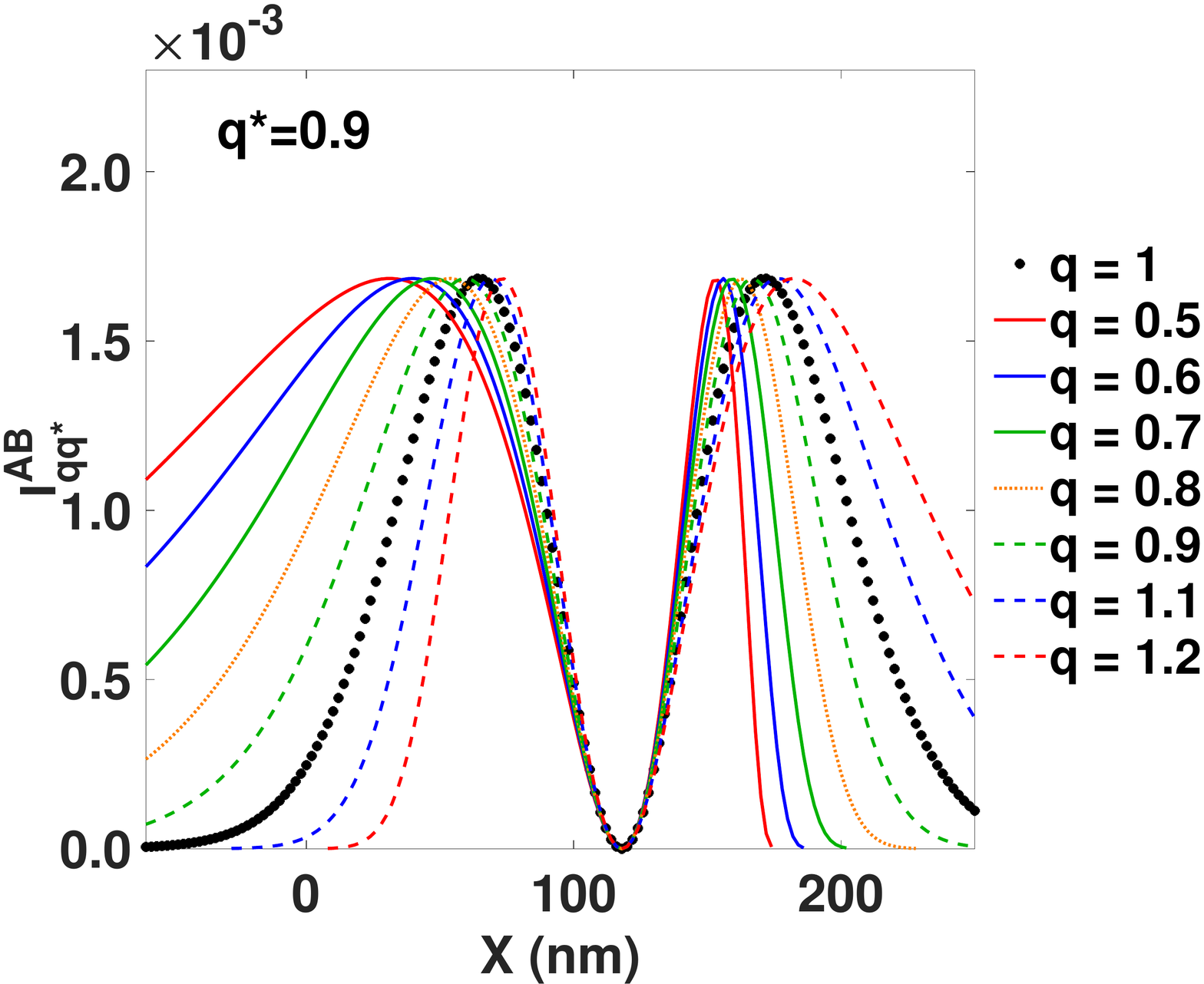}  
\end{subfigure}
\begin{subfigure}{.32\textwidth}
\centering
(b)  \includegraphics[width=0.9\linewidth]{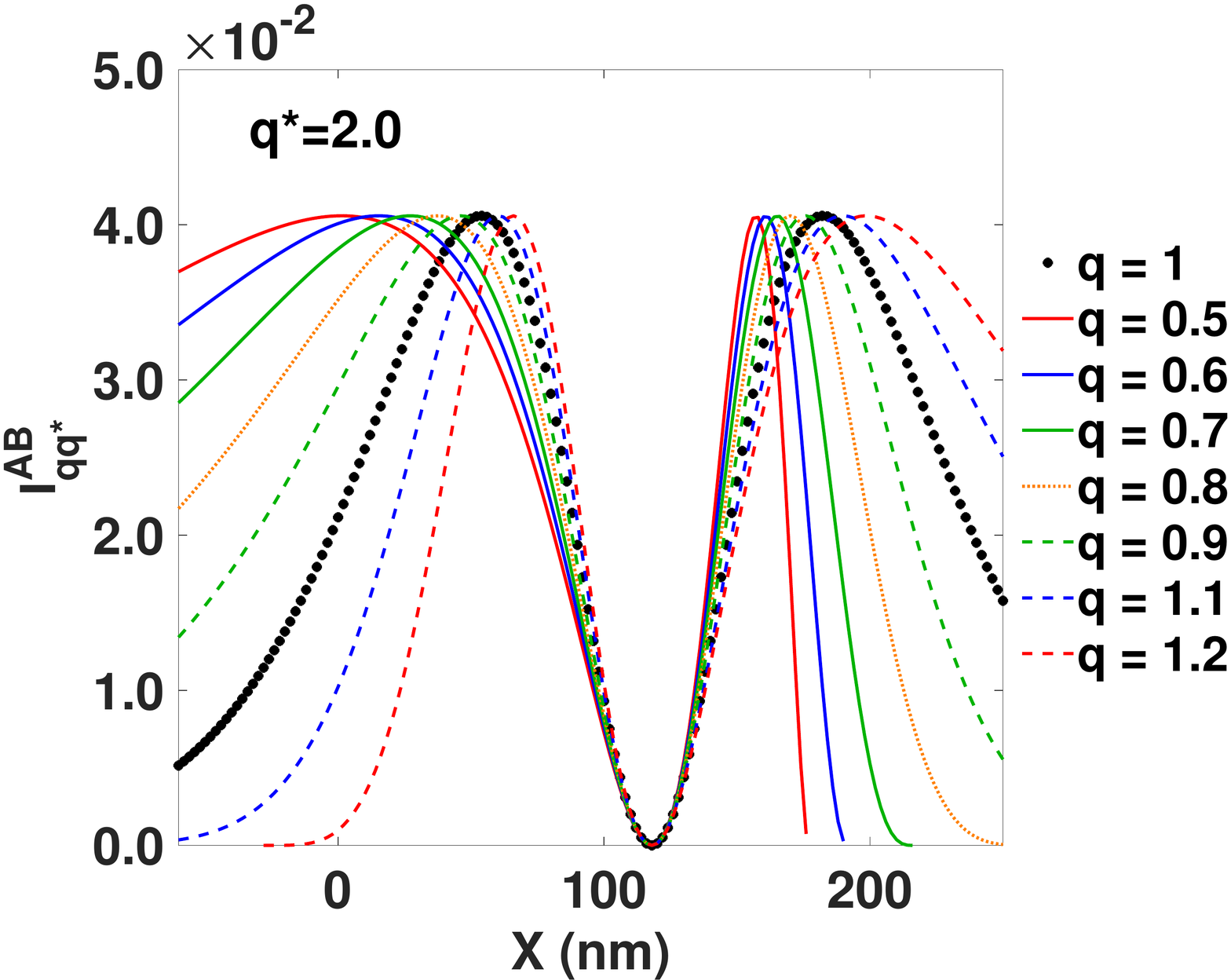}  
\end{subfigure}
\begin{subfigure}{.32\textwidth}
\centering
(c)  \includegraphics[width=0.9\linewidth]{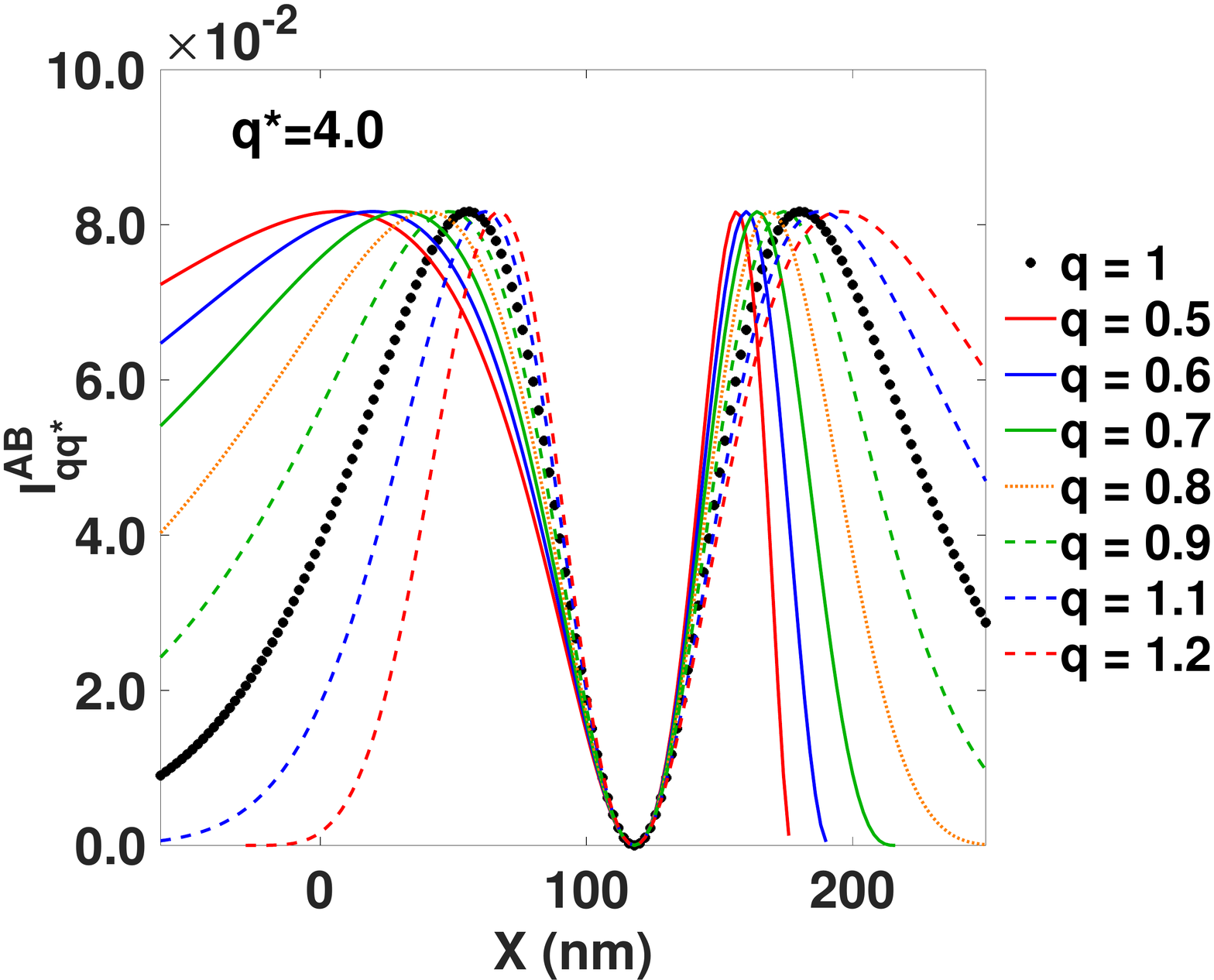}  
\end{subfigure}

\begin{subfigure}{.32\textwidth}
\centering
\vspace{0.2cm}
(d)  \includegraphics[width=0.9\linewidth]{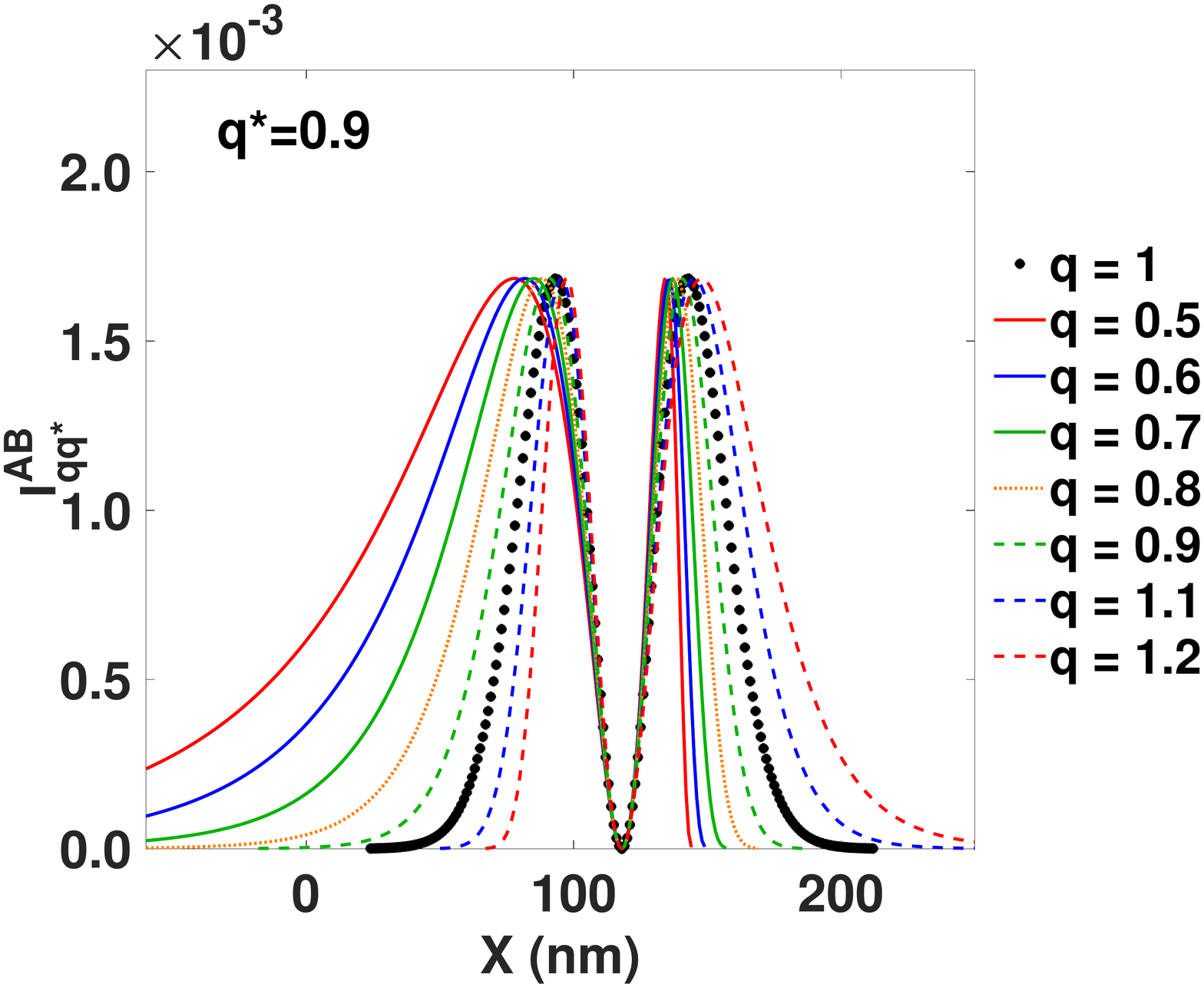}  
\end{subfigure}
\begin{subfigure}{.32\textwidth}
\centering
\vspace{0.2cm}
(e)  \includegraphics[width=0.9\linewidth]{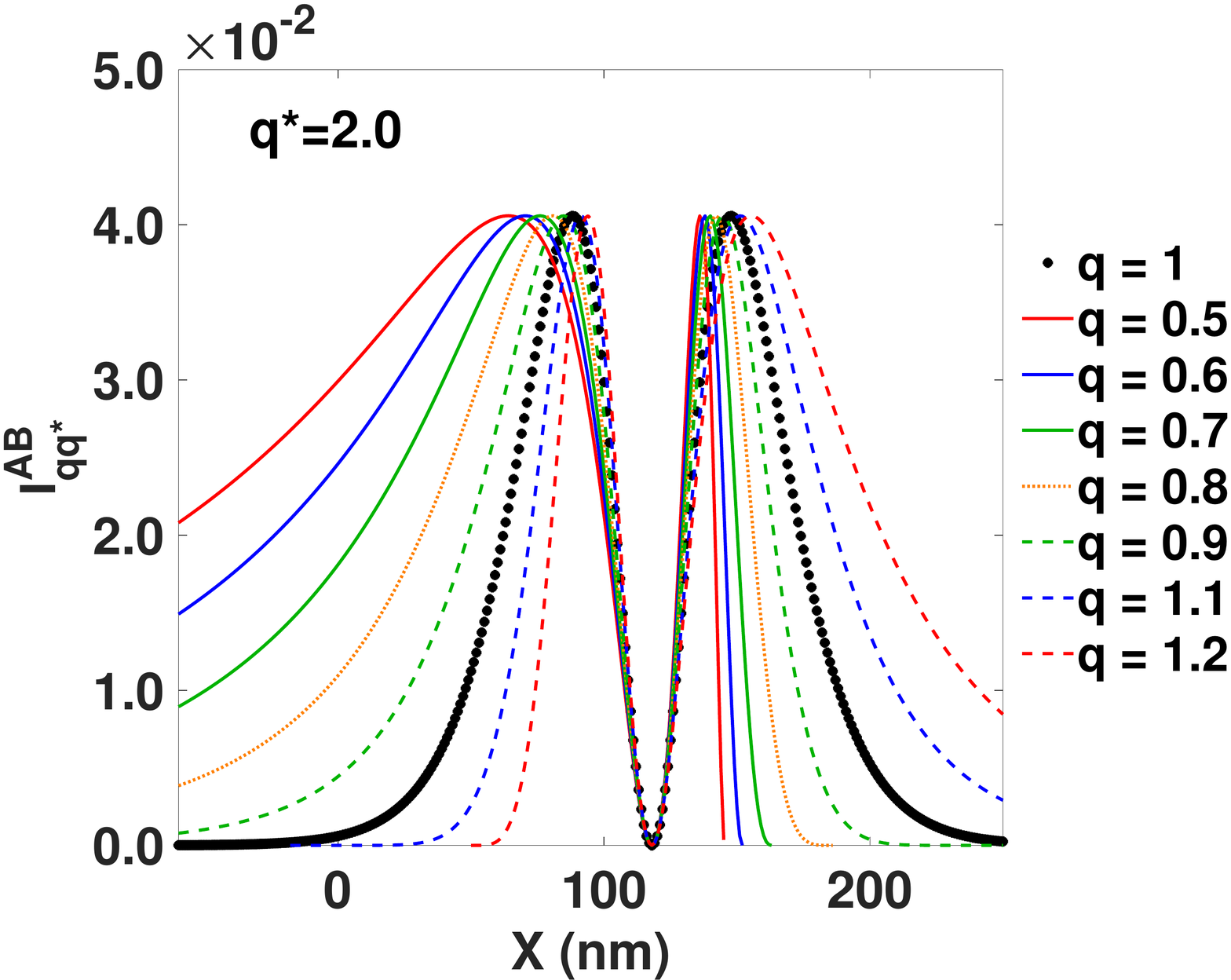}  
\end{subfigure}
\begin{subfigure}{.32\textwidth}
\centering
\vspace{0.2cm}
(f)  \includegraphics[width=0.9\linewidth]{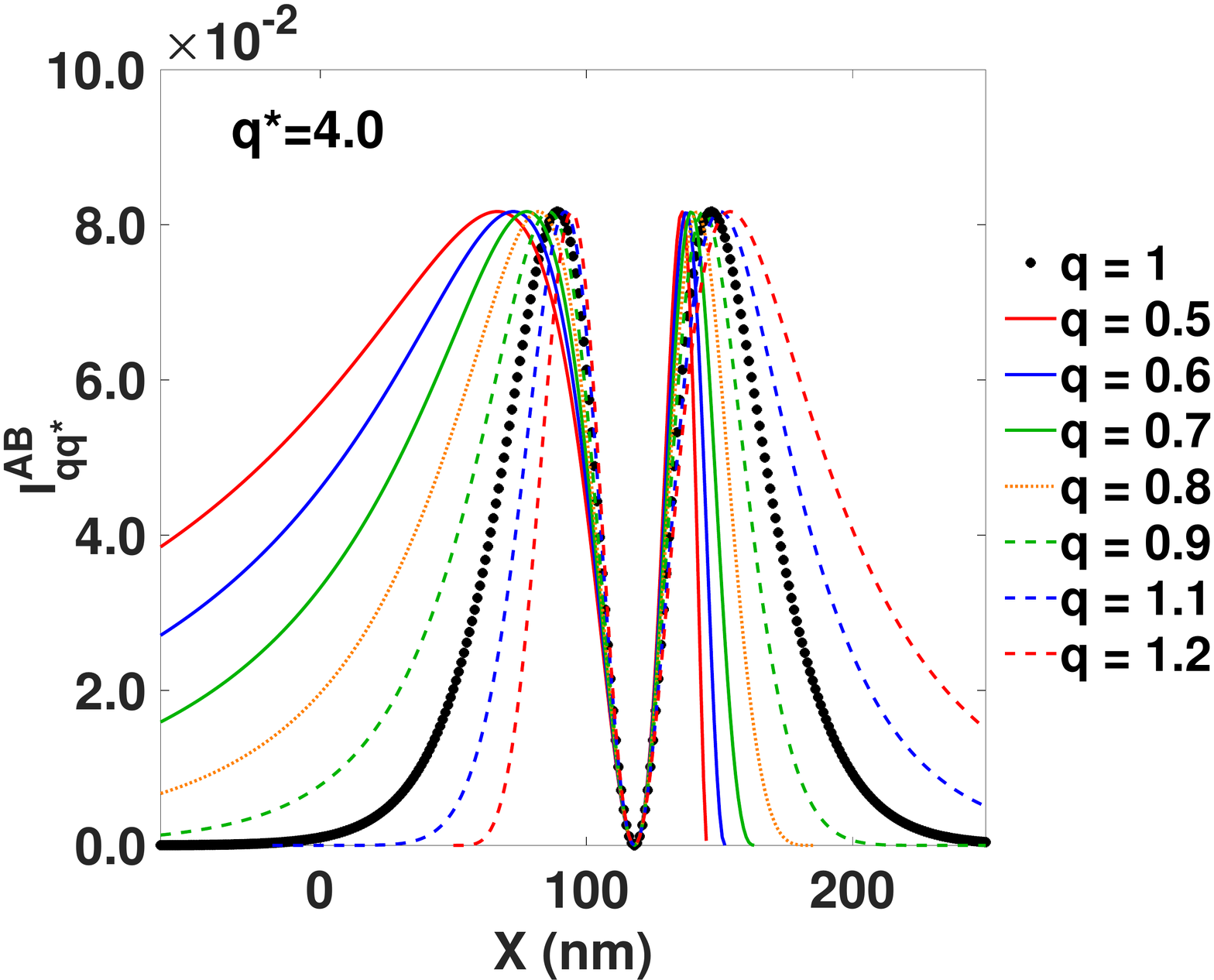}  
\end{subfigure}

\centering
\vspace{0.2cm}
\caption{Mutual information for a pair of interacting and identical $MET$ channels, as functions of the displacement for $q* = 0.9$ (a)-(d), $q* = 2.0$ (b)-(e), $q* = 4.0$ (c)-(f)  and  several $q$ values. Other channel parameters are: $z_{X}$ = $0.138$ pN is used in (a)-(b)-(c), $z_{X}$ = $0.298$ pN is assumed in (e)-(f)-(g), $X_{0}$ = $118$ nm, and \textit{T} = $21$°C extracted from (\cite{Howard1988,Hudspeth2000,Gianoli2017}).
}
\label{figInfomutua}
\end{figure}

Finally, figure \ref{figInfomutua} brings the behavior of $I_{qq*}^{AB}$, which gives a measure of the correlation between the two channels. Following the $S^{AB}_{q*}$ results, the plots also show how the magnitude of $q*$ and $z_{X}$ imposes modulations over the $I_{qq*}^{AB}$ amplitude, revealing an oscillating profile with the displacement for all $q$ values. Increasing $q*$ also exacerbates the influence of subadditivity and superadditivity, while increasing $z_{X}$ promotes a decrease in the level of $I_{qq*}^{AB}$. Additionally, there are two peaks, the first corresponding to the ascending phase of the individual probability (figure \ref{figplotpopen}) or joint activation probability (figures \ref{figplotqprodut}(a), \ref{figplotqprodut}(d), \ref{figplotqprodut}(g)), where this activation probability is above 20$\%$ and the saturation phase, in which the probability is above 80$\%$. The $I_{qq*}^{AB}$ presents a minimum at $X = X_{0}$. The first peak presents a shift to the left (direction of decreasing $X$) for a $q$ decline. The second peak shifts to the right (direction of increasing $X$) if $q$ increases. There is no shift of the minimum, changing $q$ or $q*$. The simulation also shows that higher $q*$, more accentuated, is the $I_{qq*}^{AB}$ dispersion for displacements, considering the left and right values of $X = X_{0}$. Moreover, the curves increase slowly to $X < X_{0}$ and faster for $X > X_{0}$, for lower $q$. In addition, different values of $q$, ranging from $q < 1$ to $q > 1$, modulate the $I_{qq*}^{AB}$ dispersion in response to a physical displacement of the hair bundle. In summary, our results clearly show that at $X = X_{0}$, $p_{qq*}^{OC}(X)$ and $S^{AB}_{qq*}$ reach their maximum value, while the magnitude of $I_{qq*}^{AB}$ is minimal.

\section{Discussion}

Several studies have shown that nonlinearity phenomena govern the auditory system mechanisms. Consequently, the physiology of audition arose as a pillar to investigate the presence of power laws and fractality in physiology. These reports removed the dogma that the auditory substrate was merely ruled by temporally and spatially random events. Within this scope, Eguiluz \textit{et al.} showed that a Hopf bifurcation maximizes tuning and amplification in the cochlea, explaining three origins of nonlinearities in hearing: compression of dynamic range, sharper cochlear tuning for softer sounds, and generation of combination tones (\cite{Eguiluz2000}). In addition, another letter, applying recurrence analysis to electrophysiological recordings made in the auditory system of mice, showed that the spontaneous glycinergic miniature current was not temporally random but instead exhibited a chaotic nature. Moreover, Teich presented evidence of fractality in the auditory neural spike train (\cite{Teich1994}). Further investigations also emphasized that the pattern of firing auditory nerve action potentials forms a non-renewal point process over short as well as long time scales (\cite{Lowen1992}). Recently, it was detected the presence of long-range temporal correlations in neural oscillations, considering the normal auditory system at the brainstem level (\cite{Mozaffarilegha2019}). Altogether, these findings strongly suggested that fractality and long-range correlations may be ubiquitous in hearing processing. Last but not least, Lyra and Tsallis showed that $q$ could be related to the attractor's fractal dimension at the onset of a chaotic regime. In this framework, the Boltzmann Gibbs limit $q = 1$ is associated with the case when the attractor has a Euclidean dimension, while $q > 1$ indicates the existence of a fractal attractor (\cite{Lyra1998}). Thus, in the future, it would be an essential task to verify the physical behavior of coupled $MET$ channels in the fractal realm.

Based on the discussion above, the assumption of non-extensive modeling also represents a coherent explanation in situations where long-range interactions are present. As mentioned, the bridge between non-extensivity and fractals has been systematically studied by several authors (\cite{Tsallis1995,Deppman2018}). In the nervous system, Silva \textit{et al.} evidenced that short-term plasticity, existing in the giant calyx of Held, is better modeled using a $q$-differential equation inspired by the $GTS$ formalism (\cite{Gersdorff1997,Silva2018}). Previous reports reinforce their results, showing thousands of vesicles sharing the exocytotic machinery. This morphological finding is related to the raising of interactions probability between vesicles, increasing or even disfavoring the chance of vesicular fusion in the nerve terminal (\cite{Rizzoli2005,Denker2010,Harlow2001}). In the present work, the physical interaction between ionic channels via tip-link supports a consistent groundwork for non-extensivity or long-range correlations. The generalized activation and joint probabilities configured $MET$ channel as a potential source of non-extensivity of the auditory system. This reasoning converges with Lowen and Teich's interpretation, proposing fractality in the auditory nerve as related to fractal activity in the ion channels themselves, specifically in hair cells in the cochlea (\cite{Werner2010,Lowen1993}). Also, it is convergent with a recent investigation, which demonstrated the existence of non-extensivity around an ion voltage channel pore. Besides Erden's description (\cite{Erdem2007}), these studies strengthen our physiological interpretation that an intrinsic non-extensive mechanism may rule $MET$ channels.

The generalized sigmoid function here reconciles the biomechanical arrangement originally proposed by Gianoli \textit{et al.} with the theoretical premises (\cite{Gianoli2017,Gianoli2019}). We reason that the generalized Boltzmann equation certainly is an adequate application in $IxX$ adjustments, especially in those cases where there are significant deviations from the empirical analysis and classical distribution (\cite{Holt1997,Netten2000}). Such discrepancies in this fitting procedure would be associated with the presence of non-random mechanisms in the channel kinetics as documented by Liebovitch and Sullivan, also investigated by Erden in their studies made in voltage channels (\cite{Liebovitch1987,Erdem2007}). The tip links give another possible molecular source for non-extensivity. These protein linking channels orchestrates open or closed states by producing an organized net movement of charge across both channels, reflected in the current channel response. Yet, the heterogeneity of tip-link composition might contribute to modulating the non-extensive degree (\cite{Bartsch2019}). This argument is particularly strengthened by investigations showing that tip-link exhibits variability and instability in the tension magnitude, which favors a complex behavior in the $MET$ channel coupling (\cite{Barral2011}). Lastly, $IxX$ curves depend on the hair bundle morphologies, where $MET$ channel response exhibits variations relative to specific hair bundle types (\cite{peng2011}). Altogether, these morphological features not only may regulate channel kinetics. They also may control the non-extensivity degree, reflecting on the entropy and information profiles.

The precise force-displacement relation to molecular mechanisms remains unclear. Assuming a generalized function allowed us to investigate the non-Hookean or nonlinear behavior acting on auditory hair cells. Many authors highlighted that hair bundle response is enhanced thanks to the nonlinearity signatures (\cite{Faber2018}). For instance, Russell \textit{et al.} extracted nonlinear mechanical responses of mouse cochlear hair bundles, while Kennedy \textit{et al.} used isolated cochleae preparation demonstrating a nonlinear force-displacement relationship (\cite{Russell1992,Kennedy2005}). Motivated by these findings, we scanned the pattern between the single-channel $z_{X}$ and the non-extensive $q$, enabling us to obtain a thermodynamic portrait by comparing both parameters. The two selected values for $z_{X}$  made only a slight difference in the behavior of the $p_{q,O}(X)$ curve, but it inserted a notable difference on $S_{qq*}^{AB}$ and $I_{qq*}^{AB}$ concerning $X$ magnitude. From an experimental point of view, it is known that the sigmoid shape is modulated by $z_{X}$ (\cite{Netten2000,Corey1983,Gianoli2019}). This argument for a relationship between $z_{X}$ and non-extensivity converges with the computational study by Xenakis \textit{et al.}, suggesting the non-extensive pore influence in ion channels (\cite{Xenakis2021,Beckstein2001}). Consequently, since the pore structure and gating are correlated (\cite{Beckstein2001,Aryal2015}), it is a reliable conjecture for a gating mechanism governed by $GTS$ premises. Therefore, modifications of channel kinetics reflected in $IxX$ deviations from the classical Boltzmann sigmoid function, entropic level, and informational degree arise as a reflection of the non-extensive behavior of the system. Thus, we are convinced that the non-extensive approach is more appropriate for describing the $MET$ channels thermodynamics. It takes intrinsic (non-extensive pore) and extrinsic (physical interaction via tip-link) contributions embedded into a synthetic biophysical substrate.

In their modeling of cooperation between $MET$ channels, Gianoli \textit{et al.} selected two adjacent stereocilia from the hair bundle. The scheme formulated by these authors contains two $MET$ channels on the same stereocilia connected through a tip-link to a third channel located in the neighboring stereocilia (please see figure 1 in (\cite{Gianoli2017}) for a visual inspection). In order to study the simplest case, we chose to work with two connected MET channels, leaving the inclusion of the third channel for future work. The $q*$ parameter represents the mathematical artifice to mimic the channel's binding, facilitated by the tip-link tension that regulates the channel interaction. Nevertheless, how to establish a bridge between the model and results given by Gianoli \textit{ et al.} by furnishing it with a thermodynamics context? In our particular case, how to offer a precise interpretation based on $GTS$ inspiring model? Firstly, let us suppose the absence of an external stimulus. In that case, there is no displacement ($X = X_{0}$) nor the significant influence of the tip-link over the intrinsic non-extensive channel kinetics of the pair of $MET$ channels. This configuration permits them to behave independently from each other. In other words, despite the intrinsic channel kinetics may assume a non-extensive $(q \neq 1)$ or non-extensive $(q = 1)$ behavior, the channel interaction is momentarily vanished ($q* = 0$). Secondly, the presence of an external stimulus ($q* \neq 0$) is equivalent to the shorter stereocilia moving toward the taller. Consequently, this increases the stretching of the tip link bringing channels closer by opening both of them, promoting the hair cell depolarization. Yet, in this $X$ direction, $S_{qq*}^{AB}$ will be more subadditive, especially for more positive displacements ($X > X_{0}$). Thirdly, after achieving an adaptation period, both stereocilia move in the opposite direction, weakening the tip-link strength and pulling the channels away. In this direction, the hair cell is hyperpolarized, most channels are closed, and higher $S_{qq*}^{AB}$ is associated with a superadditive regime, particularly for more negative displacements ($X < X_{0}$). It is important to emphasize that, although less reported, $q < 1$ was identified in the Bak-Sneppen model of biological systems and the turbulence of pure electron plasma experiments (\cite{Tamarit1998,Boghosian1996}). Thus, we stress that for both $X$ directions, coupled $MET$ channels may be governed by both intrinsic and extrinsic non-extensive and extensive regimes. This is afforded by the physically linked molecular mechanisms existing in stereocilia, where $q \neq 0$ and $q = 1$ are equivalent to the balance between long-range and short-range correlation processes.

Finally, it is worth mentioning that some studies on quantum correlations in spin systems have already documented an oscillatory pattern of entropy (\cite{Hazzard2014}). On the other hand, the oscillatory profile of $I_{qq*}^{AB}$ may be related to a performance compatible with a classical oscillator developed by Gianoli \textit{et al.}. The oscillating behavior extracted for $I_{qq*}^{AB}$ may be a reflection of the Markovian behavior associated with the two states used in the description of the channel dynamics, in conjunction with the spatial coupling existing between them. In this sense, works are showing that mutual information can plausibly present a time-dependent profile in the description of Markov systems (\cite{Teretenkov2020}). In summary, our analysis provides a general thermodynamic landscape, including the Boltzmann statistics, as commonly assumed by researchers of the auditory system, and added the thermodynamic pattern linked to a generalized framework.

\section{Concluding remarks}

This was the first study to provide a novel thermodynamics characterization for analyzing physically interacting $MET$ channels to the best of our knowledge. The molecular scheme here used represented a model to study the connection between a physical interaction and long-range correlation in a biophysical framework. Our results complemented the thermodynamical pattern, enlarging the knowledge acquired beyond the Boltzmann-based models. We also show the applicability of the $q$-product rule for understanding the cooperation between physically coupled ion channels. The existence of non-extensivity influences the entropy and information when there is a coupling of $MET$ channels. In general, there was the predominance of subadditivity and superadditivity regulating the entropy and mutual information depending on the displacement directions. 

Moreover, beyond observing the influence on probabilities for closing and opening the channels, we also detected the formation of oscillatory patterns in the mutual information, being these results attributed to the existence of a coupled system. Although the entropy associated with coupled oscillators has been studied in modeling several classical and quantum phenomena, it has not yet been contemplated in coupled $MET$ channels. Surprisingly, we also found that $z_{X}$ modulates the influence of non-extensive degree on joint entropy and mutual information.

In the future, we hope to expand our ideas presented here, considering investigating a system with three coupled channels. Indeed, hence we only used two $MET$ channels located in the same stereocilia it is pertinent to consider a system composed of a third $MET$ channel as well. This procedure is necessary to bring our theoretical proposal closer to the real molecular phenomenon. Yet, the expansion of the kinetic scheme here simplified will undoubtedly be the subject of future investigation. In addition, it would be important to deduce a mathematical relationship considering the single-channel $z_{X}$ as a function of the $q$. Although our investigation gives a novel theoretical view and results on how to analyze responses of hair bundles, empirical studies are welcome to verify our findings. We emphasize that the formalism presented here is not restricted to applications involving $MET$ channels. In principle, albeit our primary motivation was to study a specific mechanism of the auditory system, our formalism can be applied to investigate the existence of non-extensivity in the entropy and mutual information in any system, especially made up of two interacting channels.

\section{Acknowledgements}

We thank Dr. Riza Erden for valuable help and support during the development of this work.

\printcredits

\bibliographystyle{unsrt}

\bibliography{BIBqProduct_IonChannel_1120}

\end{document}